\documentclass{article}[11pt]

\usepackage[a4paper,top=4.0cm,bottom=2.0cm,left=2.0cm,right=2.0cm]{geometry}
\usepackage{amsmath,amsfonts,mathrsfs,amsfonts, color, bbold,bm,blindtext,ulem}
\usepackage{graphicx}% Include figure files
\usepackage{dcolumn}% Align table columns on decimal point
\usepackage{bm}% bold math
\usepackage{xcolor}
\allowdisplaybreaks

\usepackage{booktabs}
\usepackage{graphicx}% Include figure files
\usepackage{dcolumn}% Align table columns on decimal point
\usepackage{bm}% bold math
\usepackage[utf8]{inputenc}
\usepackage[T1]{fontenc}
\usepackage{mathptmx}

\usepackage{amsmath,amsfonts,mathrsfs,amsfonts}
\usepackage{epsfig,graphicx}
\usepackage{color}
\usepackage[utf8]{inputenc}
\usepackage[T1]{fontenc}
\usepackage{mathptmx}
\usepackage{cancel} 
\usepackage{algorithmicx,algorithm,algpseudocode}

\usepackage{hyperref}

\begin{document}

\title{Reconstructing regime-dependent causal relationships from observational time series}

\author{Elena Saggioro\thanks{University of Reading, Department of Mathematics and Statistics, Whiteknights, PO Box 220, Reading RG6 6AX, UK ({\tt e.saggioro@pgr.reading.ac.uk})} \and Jana de Wiljes\thanks{Universit\"at Potsdam, 
			Institut f\"ur Mathematik, Karl-Liebknecht-Str. 24/25, D-14476 Potsdam, Germany ({\tt wiljes@uni-potsdam.de})}
			\and Marlene Kretschmer\thanks{University of Reading, Department of Meteorology, Whiteknights, PO Box 220, Reading RG6 6AX, UK ({\tt m.j.a.kretschmer@reading.ac.uk})} \and Jakob Runge\thanks{German Aerospace Center, 
			Institute of Data Science, M\"alzerstr. 3, 07745 Jena, Germany ({\tt Jakob.Runge@dlr.de})}}
\maketitle

\begin{abstract}
Inferring causal relations from observational time series data is a key problem across science and engineering whenever experimental interventions are infeasible or unethical. Increasing data availability over the past decades has spurred the development of a plethora of causal discovery methods, each addressing particular challenges of this difficult task.
In this paper we focus on an important challenge that is  at the core of time series causal discovery: regime-dependent causal relations. Often dynamical systems feature transitions  depending on some, often persistent, unobserved background regime, and different regimes may exhibit different causal relations. Here, we assume a persistent and discrete regime variable leading to a finite number of regimes within which we may assume stationary causal relations. 
To detect regime-dependent causal relations, we combine the conditional independence-based PCMCI method with a regime learning optimisation approach. PCMCI allows for linear and nonlinear, high-dimensional time series causal discovery. Our method, Regime-PCMCI, is evaluated on a number of numerical experiments demonstrating that it can distinguish regimes with different causal directions, time lags, effects and sign of causal links, as well as changes in the variables' autocorrelation. Further, Regime-PCMCI is employed to observations of El Ni\~no Southern Oscillation and Indian rainfall, demonstrating skill also in real-world datasets.
\end{abstract}

\noindent
{\bf Keywords.} causal discovery, time series, non-stationarity, regime-dependence, high dimensionality, climate research  \\

%\begin{quotation}
%Regime-dependent non-stationarity is an ubiquitous feature of physical systems, especially prominent in atmospheric sciences. This dependence can be looked at as an intermittent change in relationships defining the dynamics of a multivariate system, each of which can be described as a time series causal network. In this work, we develop a novel algorithm that combines time series causal discovery for stationary network dynamics with a regime assigning linear optimisation to detect regime-dependent causal relations. Our method, Regime-PCMCI, is evaluated on a number of numerical experiments and demonstrates high performance in detecting a variety of regime-dependent features. Finally, Regime-PCMCI is applied to observations of El Ni\~no Southern Oscillation and Indian rainfall, demonstrating skill in detecting well-know seasonal regimes in a real-world dataset.
%\end{quotation}
%%%%%%%%%%%%%%%%%%%%%%%%%%%%%%%%%%%%%%%%%

\section{Introduction}

Understanding causal relationships \cite{Pearl2000,Spirtes2000} among different processes is an ubiquitous task in many scientific disciplines as well as engineering (e.g., in the context of climate research \cite{Ebert-Uphoff2012,Runge2014a,Liang2014,Kretschmer2016,HorenkoGerberetal2016,Runge2019a}, econometrics \cite{Granger1969,Droumaguetetall2016}, or molecular dynamics \cite{GerberHorenko2014}). Yet, the common approach to gaining causal knowledge by conducting experiments is often infeasible or unethical, for example in Earth sciences. All that is often given is a set of time series describing these processes with no specific knowledge about the direction and form of their causal relationships available. The challenge, termed causal discovery, is then to reconstruct the underlying graph of causal relationships from time series data \cite{Runge2019a}. Based on that graph the processes that generated the data can then be modelled in the framework of structural causal models (SCMs)\cite{Pearl2000} to further understand causal relations, predict the effect of interventions, and for forecasting.

Today's ever-growing abundance of time series datasets promises many application scenarios for data-driven causal discovery methods, but many challenges emerging from the dynamic nature of such datasets have not yet been met. Further, causal knowledge cannot be gained from data alone and each method comes with its particular set of assumptions \cite{Spirtes2000} about properties of the underlying processes and the observed data. See \cite{Runge2019a} for an overview of methodological frameworks, challenges, and application scenarios.

A particular and wide-spread challenge is regime-dependence, a common property of nonlinear dynamical systems that can also be described as one form of non-stationary behaviour. Regime-dependence means that the causal relationships between the considered processes vary depending on some prevailing background regime that may be modelled as switching between different states. Further, often such regimes have strong persistence, that is, they operate and affect causal relations on much longer time scales than the causal relations among the individual processes. 
In the climate system, for instance, several cases of such regime-dependencies exist. For example, rainfall in India in summer is known to be influenced by the so-called El Ni\~no Southern Oscillation (ENSO), an important mode of variability in the tropical Pacific affecting the large-scale atmospheric circulation and thereby weather patterns around the globe \cite{WebsterPalmer1997,ShamanTziperman2007}. It is, however, generally assumed that ENSO does only marginally affect Indian rainfall in winter \cite{Pal2015}. Thus, the causal relationships between ENSO and rainfall over India change dependent on the season that here defines the background-regime and operates on a longer time  scale (several months) than the causal relations among ENSO and Indian rainfall (several weeks).

\subsection{Existing work}\label{sec:relatedwork}
Causal discovery has seen a steep rise with a plethora of novel approaches and methods in recent years. Each approach has different underlying assumptions and targets different real world challenges as discussed in \cite{Runge2019a}. In general, causal (network) discovery methods can be classified into classical Granger causality approaches \cite{Granger1969,Barnett2015}, constraint-based causal network learning algorithms \cite{Spirtes2000}, score-based Bayesian network learning methods \cite{Koller2010,Chickering2002}, structural causal models \cite{Peters2018}, and state-space reconstruction methods \cite{Sugihara2012,Arnhold1999}. 

Here we focus on the constraint-based framework which has the advantage that it can flexibly account for nonlinear causal relations and different data-types (continuous and categorical, univariate and multivariate). PCMCI adapts this framework to the time series case yielding high detection power also in high-dimensional and strongly autocorrelated time series settings. However, one of the general assumptions of PCMCI (as well as of other causal discovery algorithms) is stationarity, i.e., that at least the existence or absence of a causal link does not change over the considered time series segment \cite{jdw:Runge2018a}. While known changes in the background signal can be accounted for by restricting the time series to the stationary regimes, PCMCI cannot handle unknown background regimes which constitute a particular case of latent confounding.

Some recent work addresses causal discovery in the presence of non-stationarity. \cite{malinsky2019learning} model non-stationarity in the form of (continuous) stochastic trends in a linear autoregressive framework.  \cite{Zhang2017c} account for non-stationarity in the more general constraint-based framework. However, both address the case of a (smoothly) varying continuous background variable that continuously changes causal relations among the observed variables. This means that these methods will not output regime-dependent causal graphs, but a ``summary'' graph that accounts for regimes modelled as latent drivers. In \cite{Peters2016a,christiansen2020switching} assumed known non-stationary regimes are exploited to estimate causal relations also in the presence of general latent confounders. 

Currently few methods exist that address the case of a discrete regime variable leading to distinct causal regimes that may be physically interpreted. For example, in the climate science context, regime-dependent autoregressive models (RAM) were introduced already in 1990 \cite{ZwiersVonStorch1990}. These can yield physically well interpretable results that, however, require well-chosen ancillary variables and a seasonal index which are not learned from data. Thus, RAM requires a priori knowledge of the regimes, which one often aims to learn rather than enforce. Furthermore, the autoregressive framework only permits linear relationships. In the context of  discrete state spaces regime dependent causal discovery has been considered in \cite{GerberHorenko2014}. Another approach that has been proposed to model time dependent Granger (non-)causality is based on a Markov Switching VAR ansatz with an economics application in mind \cite{Droumaguetetall2016}. Specifically, the regime assignments are computed by sampling from a Markov chain.

A more general framework to handle discrete regimes is the Markov-switching ansatz of \cite{jdw:deWiljesPutzigHorenko2014}, which flexibly models regime-dependence utilizing the assumption of a finite number of regimes and a level of persistency in the transitions between different regimes. This ansatz has been successfully realised in combination with many different model assumptions (e.g., see \cite{Horenko2010}) here we want to explore it for causal networks and combine it with PCMCI \cite{Runge2019b}, a constraint-based time series causal discovery method \cite{Spirtes2000}. We call our method Regime-PCMCI. 

\subsection*{}
The remainder of the paper is structured as follows: In section~\ref{sec:problemsetting} the underlying mathematical problem, concepts, and key assumptions are formalised, and a motivating example is discussed to provide some intuition. Our novel method Regime-PCMCI is then presented in section \ref{sec:method}. These  theoretical and algorithmic parts are complemented by a thorough numerical investigation of the proposed method in various artificial settings in section \ref{sec:NumInvesti}. Finally, in section \ref{sec:Climatedataex}, Regime-PCMCI is applied to a real-world dataset from climate science, addressing the changing relationships of ENSO and rainfall over India.

%%%%%%%%%%%%%%%%%%%%%%%%%%%%%%%%%%%%%%%%%%%%%%%%%%%%%%
%%%         PROBLEM SETTING                  %%%%%%%%%
%%%%%%%%%%%%%%%%%%%%%%%%%%%%%%%%%%%%%%%%%%%%%%%%%%%%%%

\section{Problem setting}\label{sec:problemsetting}
Let $\{X_t\}_{t\in \mathbb{Z}}$ be a sequence of real-valued $N_X$ dimensional random variables $X_t\in \mathbb{R}^{N_X}$ where $t$ is associated with time. A realisation over the time interval $[0,T]$ of this stochastic process is denoted $\{\mathbf{x}_t\}_{t\in[0,T]}$ and we assume that it is possible to obtain observations of these realisations. We assume that the underlying process is modelled by a regime-stationary discrete-time structural causal model (SCM)
\begin{equation}\label{eq:scm}
X^j_t = g^j_t(\mathcal{P}^j_t, \eta^j_t) \quad \text{with } j = 1, \ldots, N_X \ . 
\end{equation}
Here the measurable functions $g^j_t$ depend non-trivially on all their arguments, the noise variables $\eta^j_t$ are jointly independent and are assumed to be stationary, i.e., $\eta^j_t\sim\mathcal{D}^j$ for all $t$ for some distribution $\mathcal{D}$, and the sets $\mathcal{P}^j_t \subset (X_{t-1}, X_{t-2}, \ldots)$ define the causal parents of $X^j_t$.  Here we assume lagged relationships, but this is not a necessity. In contrast to approaches assuming stationarity, both  $g^j_t$ and $\mathcal{P}^j_t$ are allowed to depend on regimes in time as further formalized in Assumption 1 (section \ref{Ass:persistence}).
Then the problem setting considered in this manuscript is of the nature of the following inverse problem
\begin{equation}\label{eq:inverseproblem}
\mathbf{x}_t=\widehat{\mathbf{G}}_t\Big(\mathbf{x}_{t-1},\dots,\mathbf{x}_{t-\tau_{\max}}; \Theta_t\Big) 
\end{equation}
with $\widehat{\mathbf{G}}_t=[\widehat{g}^1_t,\dots,\widehat{g}^{N_X}_t]$ where $\widehat{g}^j_t$ belong to an appropriate functions space for each $t$ and $j$. $\tau_{\max}$ is the maximum considered time lag. In other words, the aim is to fit a set of unknown parameters $\Theta_t$ on the basis of an observed time series $\{\mathbf{x}_t\}_{t\in[0,T]}$. In the next section we will discuss the particular structure of the parameters $\Theta_t$ we are interested in.

\subsection{Causal graphs}
Representing causal relations between different processes as graphs (also referred to as networks) is common practice in the context of causal inference and causal discovery \cite{Pearl2000,Spirtes2000}. 
For time series, we use the concept of time series graphs. The nodes in the time series graph associated with the SCM~\eqref{eq:scm} are the individual time-dependent variables $X^j_t$ with $j = 1, \ldots, N_X$  at each time $t\in \mathbb{Z}$. Variables $X^i_{t-\tau}$ and $X^j_t$ for a time lag $\tau>0$ and a given $t$ are connected by a lag-specific directed link ``$X^i_{t-\tau} \to X^j_t$'' if $X^i_{t-\tau}\in \mathcal{P}^j_t$ for a particular $t$. We denote the maximum ground truth time lag of any parent as $\tau^{\mathcal{P}}_{\max}$.

For a more detailed introduction the reader is
 referred to \cite{Runge2019b}. In the following we will use graphs and networks interchangeably.

The collection of parent sets for all components at time $t$ is denoted $\mathcal{P}_t=\{\mathcal{P}^1_t,\dots, \mathcal{P}^{N_X}_t\}$. This set of parents is part of the unknown parameters we want to infer. Note that their dimensionality is assumed finite, but not known a priori. The other quantity of interest is the functional form of the causal relations $g^j_t(\mathcal{P}^j_t, \eta^j_t)$ in SCM~\eqref{eq:scm} corresponding to these links which we here restrict to an appropriate function class as modelled in Eq.~\eqref{eq:inverseproblem}.
If we assume linear functions with coefficients ${\Phi}^{j}_t$, then the inverse problem Eq.~\eqref{eq:inverseproblem} simplifies to 
\begin{equation}
\mathbf{x}_t=\widehat{\mathbf{G}}_t(\mathcal{P}_t;{ \Phi}_t)
\label{eq:inverse}
\end{equation}
Thus for a given time series $\mathbf{x}_t\in \mathbb{R}^{N_X}$ and with $t\in[0,T]$ and functional $\mathbf{G}_t$ the aim is to find the unknown parameters $\Theta_t=[\mathcal{P}_t,{\Phi}_t]$.

\begin{figure}[t]
    \centering
    \includegraphics[width = 0.8\linewidth]{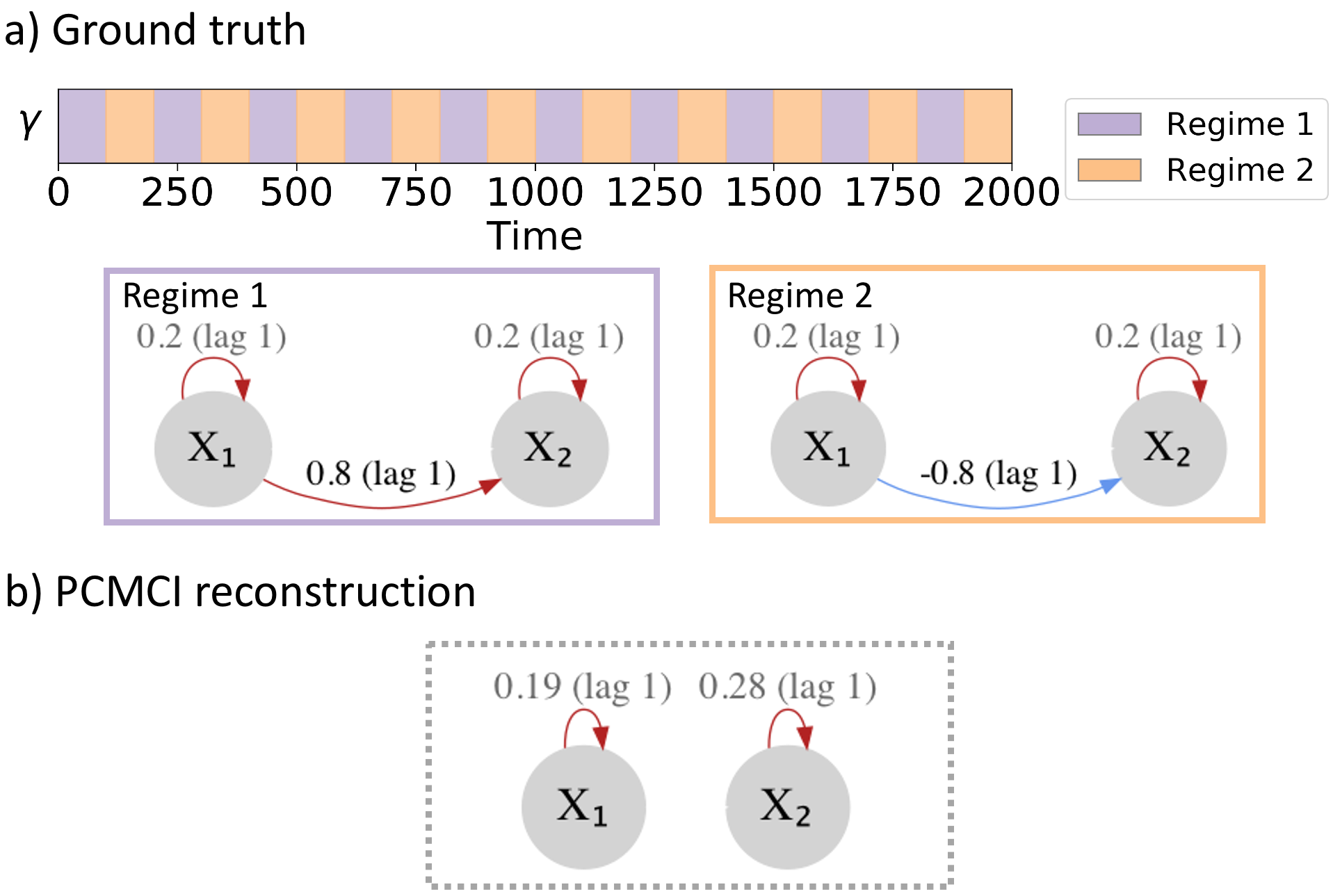}
    \caption{Motivating example. (a) Regime dependent ground truth: regime-assigning process and regime-dependent networks. The links are labelled with the associated linear coefficient $\Phi^j_k(i,\tau)$ and lag $\tau$. The sign of the coefficient is highlighted by the color (red for positive, blue for negative). (b) Network reconstruction with PCMCI estimated from the whole time series, i.e. if links are wrongly assumed to be stationary.} 
    \label{fig:wrong_signchange}
\end{figure}

\subsection{Persistence}
\label{Ass:persistence}
As mentioned above, in many application areas non-stationarity may be modelled not in form of abrupt or continuous changes, but via piece-wise constant regimes \cite{Risbeyetal2015,Williamstal2017,GerberHorenko2014}. These regimes will further exhibit a certain  persistent behaviour. In order to capture non-stationary systems with these properties we will restrict our inference to regime-dependent persistent dynamics.

\noindent\textbf{Assumption 1:}
\textit{Denote the parents and functional dependency of a given variable $j$ for a regime $k$ as $\mathcal{P}^j_t=\mathcal{P}^j_k$ and $g^j_t(\mathcal{P}^j_t, \eta^j_t)=g^j_k(\mathcal{P}^j_k, \eta^j_t)$. We call a regime $(N_M,N_K)$-persistent if the parents and functional dependencies are stationary for an average of $N_M$ consecutive time steps $t$. Further, we assume that there is a finite number of regimes on the whole time domain, i.e., $k\in\{1,\dots,N_K\}$.}
%[Persistent finite regimes]

Note that persistence enters here via a regime average persistence $N_M$, which naturally implies a finite number of regimes $N_K \leq T/N_M$.\\
Under Assumption 1 the considered linear inverse problem (\ref{eq:inverse}) reduces to finding a set of parameters 
\[
\{\mathcal{P}_1,\dots,\mathcal{P}_{N_K},\Phi_1,\dots,\Phi_{N_K}\}.
\] 
and the change points between the regimes given by the regime assigning process
\[
\Gamma(t)=[\gamma_1(t),\dots,\gamma_{N_K}(t)].
\]
with $\Gamma(t) \in [0,1]^{N_K \times T}$. For example, component $k$ of the regime assigning process can be of the form $\gamma_k=(0,1,1,0,0,0,1,\ldots) \in [0,1]^{T} $. Regime $k$ is active for all time steps for which $\gamma_k(t)=1$. 

\subsection{Motivating example}\label{sec:motexample}
Before we introduce our novel regime detecting causal discovery algorithm, we illustrate the underlying challenges of causal discovery in the face of regime-dependence by giving a simple example. 
Consider the case of two background regimes and two time series $X^1$ and $X^2$ and the associated causal graphs as shown in Figure \ref{fig:wrong_signchange},a. Variable $X^1$ linearly influences $X^2$ but the sign changes in time, alternating between a positive (during regime 1) and a negative (during regime 2) influence. Here the two regimes alternate equidistantly. 
The cross-correlation of $X^1$ and $X^2$ over the whole time-period is zero because the opposite sign effects cancel each other out in the linear regression. Thus, any linear causal discovery method would fail in detecting the influence of $X^1$ on $X^2$ when no a priori knowledge on the two background regimes exists. For example, applying a linear version of PCMCI on the whole time sample would give a network of disconnected variables (Figure \ref{fig:wrong_signchange},b). 

In contrast, if the regimes are known and PCMCI is applied to samples from both regimes separately, the positive and negative links are correctly detected (not shown). To deal with such problems automatically, our algorithm needs to learn both the regimes as well as the regime-dependent causal relations.

%%%%%%%%%%%%%%%%%%%%%%%%%%%%%%%%%%%%%%%%%%%%%%%%%%%%%%
%%%         METHOD                              %%%%%%
%%%%%%%%%%%%%%%%%%%%%%%%%%%%%%%%%%%%%%%%%%%%%%%%%%%%%%
\section{Method}\label{sec:method}
Our approach is designed to alternate between learning the regimes and the causal graphs for each regime in an iterative fashion. In principle, any causal discovery method that yields a causal graph can be used. Here we chose PCMCI \cite{Runge2019b} as a method that adapts the constraint-based causal discovery framework to the time series case. 

\subsection{Causal discovery}
The constraint-based framework has the advantage that it can flexibly account for nonlinear causal relations and different data-types (continuous and categorical, univariate and multivariate) since it is based on conditional independence defined as follows. Two variables $X$ and $Y$ are conditionally independent given a (potentially multivariate ) variable $Z$, denoted $X~{ \perp\!\!\!\perp} Y|  Z$, if
\begin{equation}
p( x,y|z)=p(x|z)p(y|z)
\end{equation}
where $p$ denotes associated probability density functions.

There exist a large variety of conditional independence tests, see \cite{jdw:Runge2018a,Runge2019b} for a discussion. If relationships are assumed linear, as is the case in practical examples of the present work, a partial correlation can be used.

As is explained in detail in \cite{Runge2019b},  PCMCI is based on a variant of the PC algorithm (names after its inventors Peter Spirtes and Clark Glymour \cite{Spirtes2000}) combined with the momentary conditional independence (MCI) test. It consists of two stages: (i)~PC$_1$ condition selection to identify relevant conditions $\widehat{\mathcal{B}}^j_t$ for all time series variables $X^j_t$ and (ii)~the MCI test to test whether $X^i_{t-\tau} \to X^j_{t}$ with 
\begin{equation} \label{eq:mit_test}
\text{MCI:}~~~~X^i_{t-\tau} ~\cancel{\perp\!\!\!\perp}~ X^j_{t} ~|~ \widehat{\mathcal{B}}^j_t\setminus \{X^i_{t-\tau}\},\,\widehat{\mathcal{B}}^i_{t-\tau}\,.
\end{equation}
Thus, MCI conditions on both the parents of $X^j_{t}$ and the time-shifted parents of $X^i_{t-\tau}$. These two stages serve the following purposes: PC$_1$ is a time-lagged causal discovery algorithm based on the PC-stable algorithm \cite{Colombo2014} that removes irrelevant lagged conditions (up to some $\tau_{\max}$) for each variable by iterative conditional independence testing. A liberal significance level $\alpha_{\rm PC}$ in the tests lets PC$_1$ adaptively converge to typically only few relevant conditions that include the causal parents with high probability, but might also include some false positives. The MCI test then addresses false positive control for the highly-interdependent time series case, which is why we chose it here.
A causal interpretation of the relationships estimated with PCMCI comes from the standard assumptions in the constraint-based framework \cite{Spirtes2000,jdw:Runge2018a,Runge2019b}, namely causal sufficiency, the Causal Markov condition, Faithfulness, non-contemporaneous effects, and  stationarity within the regimes as further discussed below. As demonstrated in \cite{Runge2019b}, PCMCI has high detection power and controlled false positives also in high-dimensional and strongly autocorrelated time series settings. 

The main free parameters of PCMCI are the chosen conditional independence test, the maximum time lag $\tau_{\max}$, and the significance levels $\alpha$ in MCI and $\alpha_{\rm PC}$ in PC$_1$. We discuss the selection of these parameters in section~\ref{sec:Modelselection}.

PCMCI is applied to sample subsets of the time series pertaining to an estimated regime $k$ in an iterative step of our method, initialised with some random regime assignment. Given a significance level $\alpha$, the output of PCMCI is the set of parents $\mathcal{P}_k=\{\mathcal{P}^1_k,\ldots,\mathcal{P}_k^{N_X}\}$ for all time series variables for that regime,
\begin{equation}
\mathcal{P}^j_k = \{X^i_{t-\tau}: pvalue_{\rm MCI}(X^i_{t-\tau},X^j_t) \leq \alpha \} \quad \forall k,j\,.
\end{equation} 
Based on these parents with associated causal links, causal effects that quantify the strength of a link can be estimated.

\subsection{Regime learning}
Given an estimated set of parents  $\mathcal{P}_k$ for each regime, the regime variable for the next iteration is updated assuming a particular non-stationary setting of finite metastable regimes as defined in Assumption 1 (section \ref{Ass:persistence}). This learning approach is based on ideas first proposed in \cite{Horenko2010} and later extended to many different models \cite{jdw:deWiljesPutzigHorenko2014}. 

In the following we focus on the linear setting, the nonlinear extension is discussed in section~\ref{sec:nonlin}. To learn the regime parameters for the inverse problem~\eqref{eq:inverse} introduced in section \ref{sec:problemsetting},
\begin{equation}
{\Phi}_k \text{ for every }k\in \{1,\dots,N_K\} \end{equation}
given $\mathcal{P}_k$ (the output from PCMCI), we define a cost functional
\begin{equation}\label{eq:costfunctional_reg}
\mathbf{L}(\Gamma,\mathcal{P})=\sum_{t=0}^{T} \sum_{k=1}^{N_K} \gamma_k(t)d({\bf x}_{t}-\mathbf{G}_t(\mathcal{P}_k; {\Phi}_k))
\end{equation}
subject to constraints 
\begin{equation}\label{eq:ConstraintGamma1}
\sum_{k=1}^{N_K}\gamma_k(t)=1\quad \forall \ t, \text{ with } \gamma_k(t) \in[0,1]
\end{equation}
and 
\begin{equation}\label{eq:ConstraintGamma2}
\sum_{t=1}^{T-1} |\gamma_k(t+1)-\gamma_k(t)|\le N_C \quad \forall k
\end{equation} 
where $d$ is a distance measure such as the squared euclidean distance $\|\cdot\|^2_2$ and $\gamma$ is a regime assigning process describing the weight of the individual networks at each time $t$. 

The format of $\mathbf{L}(\Gamma,\mathcal{P})$ relies on the assumption that the system associated with the considered data exhibits metastability in time (see Assumption 1, that translates in the summation over $k$).  Note that the persistence enters the functional in form of a regularization (see Constraint \ref{eq:ConstraintGamma2}). An alternative option is to add a regularisation term that enforces some form of smoothness of $\Gamma$ (e.g., Tikhonov regularisation \cite{Tikhonovetal1995}). 

The free tuning parameter $N_C$ is related to the average regime duration of $N_M$ time steps as follows: an average regime duration of $N_M$ in all $N_K$ regimes is implemented by choosing $N_C \approx T/(N_M N_K)$. Note that in practice, the average regime switching time $N_M$ might not be exactly known. However, we expect in many application areas that prior domain knowledge on reasonable time scales of regime switching is available. The choice of parameters, including  choices of value $N_K$, will be discussed in section~\ref{sec:Modelselection}.

%%%%%%%%%%%%%%%%%%%%%%%%%%%%%%%%%%%%%%%%%%%%%%%%%
%%%         Pseudocode                    %%%%%%
%%%%%%%%%%%%%%%%%%%%%%%%%%%%%%%%%%%%%%%%%%%%%%%%%

\subsection{Algorithm}
The Regime-PCMCI algorithm iterates over two major estimation steps: (Step 1) causal discovery to obtain $\mathcal{P}_k$ and fit the coefficients $\Phi_k$ (or a nonlinear function) and (Step 2) regime learning to update the regime variable $\Gamma$. In the following, $q$ indicates the current iteration. The superscript $(q)$ is added combined with brackets to the variables updated in each loop. The details of the consecutive subroutines are laid out below.

\subsubsection{Step 1: Causal discovery and model estimation}

The first step is to estimate a set of parents $\{\mathcal{P}_k\}^{(q)}$ and coefficients $\{\Phi_k\}^{(q)}$ with $k\in\{1,\dots,N_K\}$ on the basis of a fixed $\{\Gamma(t)\}^{(q)}$ obtained in step 2 of the previous iteration (see lines of Algorithm \ref{alg:ESRF} and section \ref{sect:step2}). In the first iteration, the regimes are assigned randomly.  $\{\mathcal{P}_k\}^{(q)}$ and $\{\Phi_k\}^{(q)}$ are estimated on the basis of a subset of the time series ${\bf x}_{t}$ with \begin{equation}
t \in \{\Upsilon_k\}^{(q)} := \Big\{ t \, : \, \{\gamma_{k}(t)\}^{(q)} \geq 0.5 \Big\} 
\end{equation}
for each regime $k$. The regime-dependent parents $\{\mathcal{P}_k\}^{(q)}$ are estimated via PCMCI. 

To solve Eq.~\eqref{eq:inverse}, we assume a functional relationship ${\bf G}$ that relates each variable to its parents $\mathcal{P}_k$ (for each regime). Here we assume linear functions $g^j_k$ implying that the coefficients $\Phi_k$  can be estimated from the following regression model for each fixed $k$:
 \begin{equation}
x^j_t =\sum_{X^i_{t-\tau}\in \mathcal{P}^{j, (q)}_{k}} \{\Phi^j_k(i,\tau)\}^{(q)} x^i_{t-\tau} + \varepsilon^j_t 
\end{equation}
for $t \in\{ \Upsilon_k\}^{(q)}$. In other words for every $k\in\{1,\dots,N_K\}$ the following optimisation has to be solved 
\begin{equation}\label{eq:linearregression}
\{{ \Phi}^j_k(i,\tau)\}^{(q)}=\arg \min \Big\|x^j_t - \sum_{X^i_{t-\tau}\in \mathcal{P}^{j,(q)}_{k}} \{{\Phi}^j_k(i,\tau)\}^{} x^i_{t-\tau}  \Big\|^2_2 
\end{equation} 
for $t \in\{ \Upsilon_k\}^{(q)}$.
Note that the coefficients not indicated as relevant via the parent set are defined to be zero, i.e., ${ \Phi}^j_k(i,\tau):=0$ for $X^i_{t-\tau}\notin \mathcal{P}^{j,(q)}_{k}$.

\subsubsection{Step 2: Regime learning}\label{sect:step2}
Step 2 is to determine an optimal regime assigning process $\{\Gamma_t\}^{(q+1)} \in [0,1]^{N_K \times T}$ given the current estimates  $\{\mathcal{P}_k\}^{(q)}$ for the parents and $\{\Phi_k\}^{(q)}$ coefficients (see second bullet in $q$-loop of Algorithm \ref{alg:ESRF}). For this the following optimisation problem needs to be solved
  \begin{equation}\label{eq:Gammaoptimization}
\{\Gamma_t\}^{(q+1)}=\arg \min \sum^{N_K}_{k=1}\sum^{T}_{t=1} \gamma_k(t) \Big\|{\bf x}_t-  \{{\hat{\bf x}_{k,t}\}^{(q)}}\Big\|^2_2
\end{equation}
subject to the constraints (\ref{eq:ConstraintGamma1}) and (\ref{eq:ConstraintGamma2}), and where for each $k \in \{1,\dots,N_K\}$
\begin{equation} \label{eq:x_hat_k}
 \hat x^j_{k,t} = \sum_{X^i_{t-\tau}\in \mathcal{P}^{j}_{k}} {\Phi}^j_k(i,\tau) x^i_{k,t-\tau} \quad \text{ for }  t \in\{1,\dots,T\}.
\end{equation}
Since the first $\tau_{\max}$ time steps cannot be predicted, we choose to set those to $\hat x^j_{k,t} =x^j_{k,t} $ and to not consider this portion of the time series in the algorithm evaluation. 

In order to search for the global minimum, the algorithm is run for a number $N_A$ of different initializations of $\{\Gamma\}^{(0)}$ (annealing). The annealing run with the lowest cost functional objective is chosen as optimal fit. Note that the individual annealing steps are \textit{embarrassingly parallelizable}.

\begin{algorithm}[]
\caption{Regime-PCMCI}\label{alg:ESRF}
\begin{algorithmic}
\State {\textbf{Input:}
\begin{itemize}\item Time series ${\bf x}_{t}\in \mathbb{R}^{N_X}$ with $t\in\{1,\dots,T\}$ \item Parameters:\begin{itemize}   
\item Number of assumed regimes $N_K$
\item Maximum number of transitions within a single regime $N_C$
\item []
\item Maximum time lag $\tau_{\max}$
\item Functional model $\bf G$
\item Conditional independence test according to $\bf G$ (e.g. partial correlation for linear $\bf G$) 
\item Significance level $\alpha$ (and $\alpha_{\rm PC}$ for PC$_1$ step)
% \item type of masking ‘$y$' 
\item []
\item Annealing steps $N_A$
\item Number of optimisation iterations $N_Q$
    \end{itemize}\end{itemize}}
\\
\For {$a=0:N_A$} 
\State {Initialize random $\{\Gamma\}^{(0)}\in[0,1]^{N_K\times T}$ }
	\For {$q=0:N_Q$} 

	\State{
	   \textit{Causal discovery and model estimation:}
\begin{itemize}
			 %\item Assign each $t$ to a single regime according to $\Gamma_{(q)}$ : $t \in \Upsilon_k$ if $ \gamma^k_{(q)}(t) \geq \gamma^*>0.5$ ;
			\item Infer parents $\{\mathcal{P}_k\}^{(q)}$ by means of PCMCI run on subset $\Big\{{\bf x}_t:t \in\{ \Upsilon_k\}^{(q)}\Big\}$ for each $k$
			\item Fit model coefficients $\{\Phi_k\}^{(q)}$ via \eqref{eq:linearregression} for each $k$, and use them to
			generate $k$ reconstructed time series $\{\hat {\bf  x}_{k,t}\}^{(q)}$ defined for every $t \in \{1,\dots,T\}$ according to \eqref{eq:x_hat_k}. 
		\end{itemize}}
	\State{	
	\textit{Fit regime assigning process:} 
	\begin{itemize}
	\item Update $\{\Gamma\}^{(q+1)}$ solving \eqref{eq:Gammaoptimization}.
	\end{itemize}}
\State {Break if $\{\Gamma\}^{(q+1)} = \{\Gamma\}^{(q)}$ (a local or global minimum is reached)}
\EndFor
\EndFor
\\
\State {\textbf{Output:}\\
\begin{itemize}
	\item $\Gamma=[\gamma_1(t),\dots,\gamma_{N_K}(t)]^{\dag}\in[0,1]^{N_K\times T}$ 
\item Causal parents $\mathcal{P}_k$ and causal effects $\Phi_k$ for every $k\in \{1,\dots,N_K\}$ 
\end{itemize}}
\Return 
\end{algorithmic}
\end{algorithm}

\subsection{Reconstruction of time series}
A single prediction from Eq.~\eqref{eq:x_hat_k} can be derived as the weighted sum over $k$ 
\begin{equation} \label{eq:x_hat}
 \hat x^{*j}_t = \sum_{k=1}^{N_K} \lceil\gamma_k(t)\rceil \hat x^j_{k,t} \quad \text{ for }  t \in\{1,\dots,T\}.
\end{equation} But note this is never used in the code (only \eqref{eq:x_hat_k} via its presence in \eqref{eq:Gammaoptimization} is used).

\subsection{Nonlinear functions}
\label{sec:nonlin}
It is important to mention that the choice of functions $g^j_k$ in the learning problem~\eqref{eq:inverseproblem} should be determined according to the considered applications and on assumptions on the data. Further, the conditional independence test used in PCMCI should cover at least an equally expressive functional dependency class. For example, if Gaussian processes are used to estimate $g^j_k$, then the Gaussian Process Distance Correlation (GPDC) test (see \cite{Runge2019b}) can be used.

Consequently, a nonlinear version of the presented Regime-PCMCI would require a different cost functional. 
The complexity of the assumed model would increase significantly due to the two-fold presence of non-linearity (one through the regime-dependence and the other one via nonlinear causal relations). Therefore, we here restricted the focus to linear functions $g^j_k$. Addressing  nonlinearity in combination with the considered non-stationarity will be explored in subsequent research.

\subsection{Parameter selection}\label{sec:Modelselection}
Regime-PCMCI involves a number of parameters that need to be chosen. They can be separated into parameters of the causal discovery method PCMCI and those of the regime learning part.

The main free parameters of PCMCI are the chosen conditional independence test, the maximum time lag $\tau_{\max}$, and the significance levels $\alpha$ in MCI and $\alpha_{\rm PC}$ in PC$_1$. $\alpha_{\rm PC}$ should be regarded as a hyper-parameter and can be chosen based on model-selection criteria such as the Akaike Information Criterion (AIC) \cite{Akaike1973} or cross-validation. $\tau_{\max}$ could be incorporated into this  model selection. But since PCMCI is not very sensitive to this parameter \cite{Runge2019b} (as opposed to, e.g., Granger causality), its choice can be based on lagged correlation functions, see \cite{Runge2019b} for a discussion. The choice of conditional independence test is a modelling assumption guided by the assumed nonlinearity of the underlying process and also finite sample considerations. Finally, $\alpha$ is chosen based on the desired level of false positives.

Determining a suitable choice of the unknown number of regimes $N_K$ is a difficult task. In particular, it is hard to find the right balance between avoiding to overfit and to choose appropriately complex models to describe a specific dataset and thus the underlying dynamics well. One way to assess this balance heuristically is to employ an information criterion (IC) \cite{BurnhamAnderson2002} which has been derived in the context of regression models and since been adapted to various other model scenarios including graphs \cite{ShipleyDouma2019}.

An IC is designed to capture the goodness of fit penalised by the number of parameters in order to prefer models with as few parameters as possibles, to avoid overfitting (parsimony). Here the number of parameters is defined as
\begin{equation}\label{eq:numberofparameters}
N_{\text{para}}=(N_K-1)N_C+\sum^{N_K}_{k=1}\sum^{N_X}_{j=1} |\mathcal{P}^j_k|.
\end{equation}
The first term in Eq.~\eqref{eq:numberofparameters} relates to the number of parameters required to describe $\Gamma$ which can be fully determined via the change points. The second term in Eq.~\eqref{eq:numberofparameters} counts the number of relevant parents, that is, the non-zero coefficients ${\Phi}^j_k(i,\tau)$. 
Here we  use the corrected Akaike Information criterion (AICc) first proposed in \cite{HurvichTsai1989} to estimate $N_K$. Note that we use the corrected version of the original AIC \cite{Akaike1973} to correct for small samples sizes relative to the number of parameters
\begin{equation}\label{eq:AICc}
AICc =- 2 \log(\mathcal{L})+ 2 N_{\text{para}} +\frac{2N_{\text{para}}(N_{\text{para}}+1)}{N_T-N_{\text{para}}-1}
\end{equation}
where $\mathcal{L}$ is the maximum value of the likelihood function for the model one assumes for the residuals (see \cite{Metzneretal2012} for a more detailed discussion). The choice of $N_K$ is numerically investigated in section~\ref{sec:K_selection}. 

The number of iteration steps $N_Q$ should be chosen to ensure that the optimisation process converges. In our experiments we found with exploratory testings that $N_Q$ shows convergence after about 10-20 iterations for all examples investigated. The number of annealing steps $N_A$ should be chosen to ensure we can span a large number of local solutions to this non-convex optimisation problem (Eq.~\eqref{eq:costfunctional_reg}). However, computational time will set a limit to a too high parameter. Note, however, that this part is embarrassingly parallelisable.

%%%%%%%%%%%%%%%%%%%%%%%%%%%%%%%%%%%%%%%%%%%%%%%%%%%%%%
%%%         Numerical Investigation             %%%%%%
%%%%%%%%%%%%%%%%%%%%%%%%%%%%%%%%%%%%%%%%%%%%%%%%%%%%%%

\section{Numerical investigation}\label{sec:NumInvesti}
In the following we investigate the performance of Regime-PCMCI by means of several toy examples. The artificial data is designed to test the methods robustness and accuracy with respect to various potential scenarios that could occur in real applications. At first low dimensional ($N_X=2$) causal relations are studied as the results can be interpreted more easily. Next, we also consider higher dimensional settings ($N_X=10$).
The reference time series are generated with the following SCM time series model:
\begin{equation}%\label{eq:toy_timeseries}
\begin{aligned} \label{eq:toy_timeseries}
x^j_t = \sum_{k=1}^{K}&\{\gamma_k(t)\}^{\text{ref}} \sum_{X^i_{t-\tau}\in \mathcal{P}^{j}_{k}} \{{ \Phi}^j_k(i,\tau)\}^{\text{ref}} x^i_{t-\tau} + {\bf \varepsilon}^j_t, \\
&\varepsilon^j_t \sim \mathcal{N}(0,\{\sigma^2\}^{\text{ref}})
\end{aligned}   
\end{equation}
%\jr{Isn't this the way we compute $\hat{\bf x}$ (without the noise)? Maybe use this instead of Eq. (15)?}\es{ Elena: Jakob, No it is not. The reconstructed time series in Eq (15), $x_k,t$, has $N_K$ values for each $t$ (needed to update a $\Gamma \in [0,1]^{N_K \times T}$) while Eq. (18) only has one value for each $t$. Does it make sense?}
with predefined $\{\Gamma(t)\}^{\text{ref}}$, $ \{{ \Phi}_k\}^{\text{ref}}$, and $\{\sigma^2\}^{\text{ref}}$. Note that the reference set of parents is specified by the non-zero coefficients $ \{\Phi^j_k(i,\tau)\}^{\text{ref}}$.

%%%%%%%%%%%%%%%%%%%%%%%%%%%%%%%%%%%%%%%%%%%%%%%%%
%%%%%%%%%%%%%%%%%%%%%%%%%%%%%%%%%%%%%%%%%%%%%%%%%

\subsection{Low dimensional data with two underlying regimes}\label{sec:Lowdim_linear} First we focus on a  simple setting of two regimes, i.e. $\{N_K\}^{\rm ref}=2$, and a two dimensional underlying process $X_t\in\mathbb{R}^2$ (i.e., $N_X=2$). Our aim is to test the performance of Regime-PCMCI for different elemental features that can change between regimes. For brevity, links $X^i_{t-\tau} \to X^j_{t}$ will be called auto-links or auto-dependencies for $i=j$ and cross links for $i\neq j$. We consider the following scenarios as summarised in Table \ref{tab:linearexp}: sign change of coefficient (in auto link and cross variables link), lag change (in cross link), coefficient  change (in auto link) and child-parent inversion defined via an assortment of linear functions and associated coefficients. In all examples, each variable is also auto-linked at lag 1, which is a realistic yet challenging assumption for many algorithms. 
% First, we  describe the specific design of the synthetic datasets and the methodological settings. 

\subsubsection{Experiment settings}
We design five toy models, in network terms, corresponding to different sets of parents defined via the references parameters $\{{\Phi}^j_k(i,\tau)\}^{\text{ref}}$ given in columns $4$ to $5$ of Table \ref{tab:linearexp}. Further, synthetic regime assigning processes $\{\Gamma(t)\}^{\text{ref}}$ are generated for all examples. More specifically, $\{\gamma_1(t)\}^{\text{ref}}$ is designed to consist of $41$ alternating windows, i.e., $\{N_C\}^{\text{ref}}=40$ regime transitions.
% The regime assignment is indicated by setting it to $1$ (active regime) and $0$ (inactive regime). 
The length of these windows is randomly selected to be between $70$ and $100$ and the constraint (\ref{eq:ConstraintGamma1}) imposes $\{\gamma_2(t)\}^{\text{ref}}=1-\{\gamma_1(t)\}^{\text{ref}}$. The final length of the time series is capped at $T=3,000$ to ensure equally-long regime assignment time series.

Then an artificial time series ${\bf x}_t$ via (\ref{eq:toy_timeseries}) with $\{\sigma^2\}^{\text{ref}} = 1$ is generated. Note that the stochastic process (\ref{eq:toy_timeseries}) can be exactly reconstructed via the coefficients $\{{\Phi}^j_k(i,\tau)\}^{\text{ref}}$, their activation $\{\Gamma(t)\}^{\text{ref}}$ and a specific realisation of the innovation term ${\bf \varepsilon}^j_t $. 

The PCMCI parameters are chosen as follows: partial correlation as a conditional independence test, $\alpha=0.01$, $\alpha_{\rm PC} =0.2$ as recommended in \cite{Runge2012b}, $\tau_{\max}=3$, and masking type ‘y' (see the documentation of \texttt{tigramite} for the definition of masking types). %and the option to correct for a network-wide false discovery rate is turned off.
The number of regimes was set to $N_K=2$ and the maximum number of regime transitions is $N_C=40$, i.e., correct guess on number of regimes and switches (model selection for $N_K$ is investigated in section~\ref{sec:K_selection}). The number of iterations is $N_Q=20$ and the number of annealings is $N_A = 50$. A summary of the parameters is shown in Table \ref{tab:linearlowdimruns_settings}.
% The reconstructed time series is generated via Eq.~\eqref{eq:x_hat} \jr{Double check...} where coefficients $\{ \Phi_{k} \}$ are estimated for each variable via multiple linear regression on the detected parents. \\
We generate $N_{R}=100$ time series realisations for each example.

% table entries match examples
\begin{table*}[!ht  ]\center
\setlength{\tabcolsep}{8pt} % Default value: 6pt
\renewcommand{\arraystretch}{1.5} % Default value:1 
 \begin{tabular}{ l  |c | c | c| c}%c r }
%\hline
    \textbf{Example} & $k=1$ &$k=2$ & $\{{\Phi}^j_1(i,\tau)\}^{\text{ref}}$ & $\{{\Phi}^j_2(i,\tau)\}^{\text{ref}}$ \\% &$\{N_c\}^{\text{ref}}$ &$\{\sigma^2\}^{\text{ref}}$\\
  \hline
      \textit{arrow direction}  &$X^1 \rightarrow X^2$  &$X^1 \leftarrow X^2$ & $\{{\Phi}^2_1(1,1)\}^{\text{ref}}=0.8$ &$\{{\Phi}^1_2(2,1)\}^{\text{ref}}=0.8$\\%\\ &40 &$\mathbb{Id}$\\  
    & & & $\{{\Phi}^1_1(1,1)\}^{\text{ref}}=0.2$&$\{{\Phi}^1_2(1,1)\}^{\text{ref}}=0.2$ \\%&&\\
        & & & $\{{\Phi}^2_1(2,1)\}^{\text{ref}}=0.2$&$\{{\Phi}^2_2(2,1)\}^{\text{ref}}=0.2$ \\%&&\\
\hline
    \textit{causal effect} & $X^1 \xrightarrow{|a|} X^1$ & $X^1 \xrightarrow{|b|} X^1$& $\{{\Phi}^1_1(1,1)\}^{\text{ref}}=0.8$& $\{{\Phi}^1_2(1,1)\}^{\text{ref}}=0.1$ \\%&40 &$\mathbb{Id}$\\
    & & & $\{{\Phi}^2_1(2,1)\}^{\text{ref}}=0.4$& $\{{\Phi}^2_2(2,1)\}^{\text{ref}}=0.4$\\% & &\\
\hline
    \textit{lag}  & $X^1 \xrightarrow{\tau=1} X^2$ & $X^1 \xrightarrow{\tau=2} X^2$& $\{{\Phi}^2_1(1,1)\}^{\text{ref}}=0.8$& $\{{\Phi}^2_2(1,2)\}^{\text{ref}}=0.8$\\  %&40&$\mathbb{Id}$ \\
    & & & $\{{\Phi}^1_1(1,1)\}^{\text{ref}}=0.2$& $\{{\Phi}^1_2(1,1)\}^{\text{ref}}=0.2$\\% && \\
      & & & $\{{\Phi}^2_1(2,1)\}^{\text{ref}}=0.2$& $\{{\Phi}^2_2(2,1)\}^{\text{ref}}=0.2$ \\%&& \\
\hline
 % \hline
   \textit{sign $X^1$}  & $X^1 \xrightarrow{|a|} X^1$ & $X^ 1\xrightarrow{-|a|} X^1$ & $\{{\Phi}^1_1(1,1)\}^{\text{ref}}=0.8$
    & $\{{\Phi}^1_2(1,1)\}^{\text{ref}}=-0.8$ \\% &40& $\mathbb{Id}$ \\
& & & $\{{\Phi}^2_1(2,1)\}^{\text{ref}}=0.2$& $\{{\Phi}^2_2(2,1)\}^{\text{ref}}=0.2$\\% & &\\
\hline
    \textit{sign $X^1X^2$}  &$X^1 \xrightarrow{|a|} X^2$ & $X^1 \xrightarrow{-|a|} X^2$& $\{{\Phi}^2_1(1,1)\}^{\text{ref}}=0.8$ & $\{{\Phi}^2_2(1,1)\}^{\text{ref}}=-0.8$ \\%&40 &$\mathbb{Id}$ \\
    & & & $\{{\Phi}^1_1(1,1)\}^{\text{ref}}=0.2$&  $\{{\Phi}^1_2(1,1)\}^{\text{ref}}=0.2$ \\%&& \\
        & & & $\{{\Phi}^2_1(2,1)\}^{\text{ref}}=0.2$&  $\{{\Phi}^2_2(2,1)\}^{\text{ref}}=0.2$ \\%&& \\

\hline
  \end{tabular}
   \caption{Artificial model configurations for different low dimensional experiments with $N_K=2$ underlying regimes.}
   \label{tab:linearexp}
\end{table*}

\begin{table}
\setlength{\tabcolsep}{8pt} % Default value: 6pt
\renewcommand{\arraystretch}{1.5} % Default value:1 

\begin{tabular}{ c  c  c c | c  c c c}
%\hline
  CI test & $\tau_{\max}$ &  $\alpha$ & $\alpha_{\rm PC}$ & $N_K$ & $N_C$ & $N_Q$ & $N_A$\\ 
\hline
%\hline
    ParCorr  & $3$   & $0.01$     &$0.2$     & $2$        & $40$ & $20$  & $50$   \\
%\hline
  \end{tabular}
  \caption{Method parameters for low dimensional examples with $N_K=2$ underlying regimes.}
   \label{tab:linearlowdimruns_settings}
\end{table}

\subsubsection{Results}
\begin{figure}
\center
\includegraphics[width=0.8\linewidth]{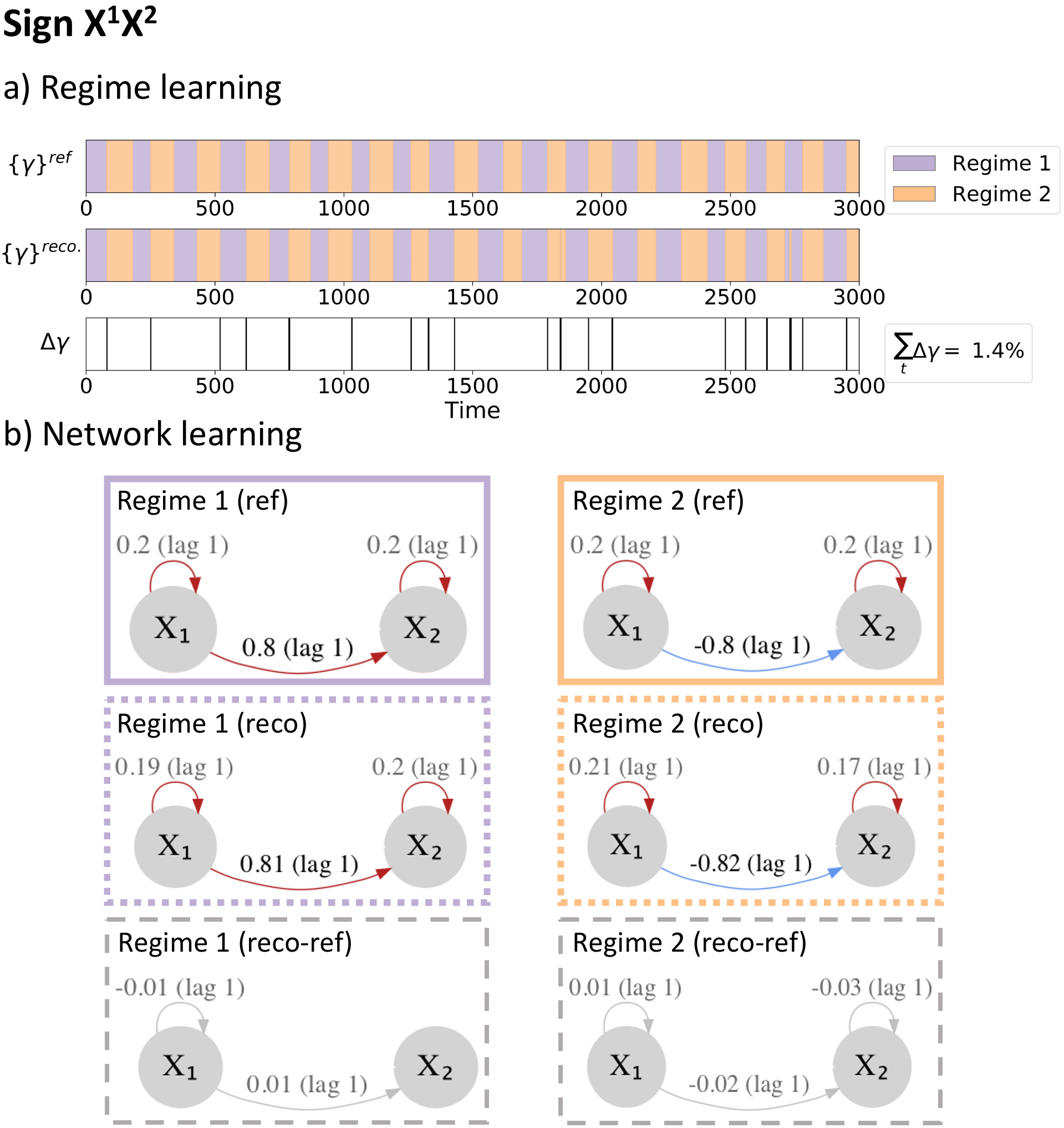}
\caption{Example case \textit{Sign $X^1X^2$}. (a) The ground-truth regime-assigning process, $\{\gamma\}^{ref}$ (top), the Regime-PCMCI reconstructed process, $\{\gamma\}^{reco.}$ (middle) and the difference between the two, $\Delta \gamma$ (bottom). (b) The ground-truth networks for each regime (top), the Regime-PCMCI reconstructed networks (middle) and the difference between the two (bottom). The links are labelled with the associated linear coefficient $\Phi^j_k(i,\tau)$ and the lag $\tau$. The sign of the coefficient is highlighted by the color (red for positive, blue for negative).}
\label{fig:signXY}
\end{figure}

\begin{figure}
\center
\includegraphics[width=0.8\linewidth]{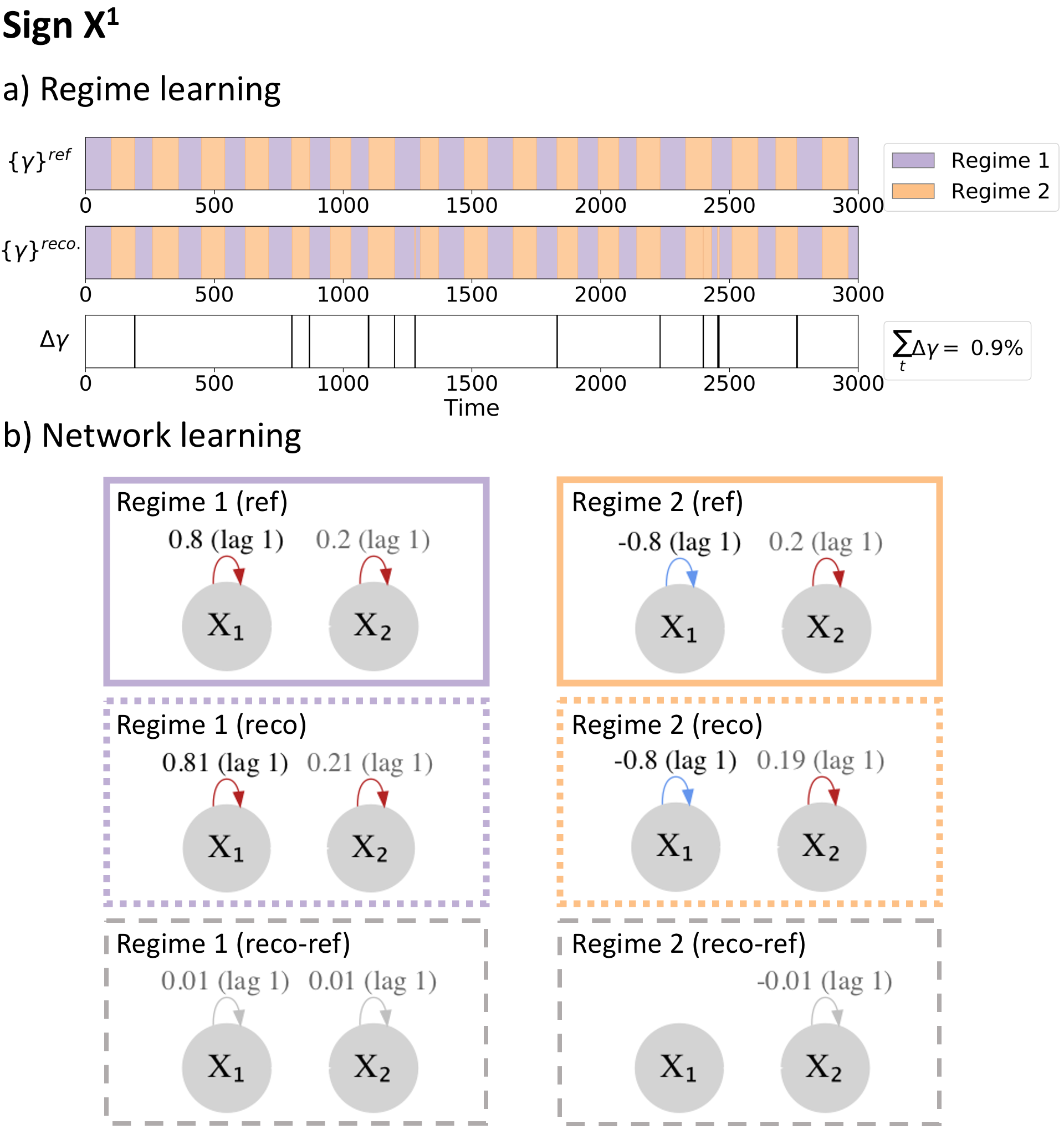}
\caption{Example case \textit{Sign $X^1$}. See description in Figure \ref{fig:signXY}. }
\label{fig:signX}
\end{figure}

\begin{figure}
\center
\includegraphics[width=0.8\linewidth]{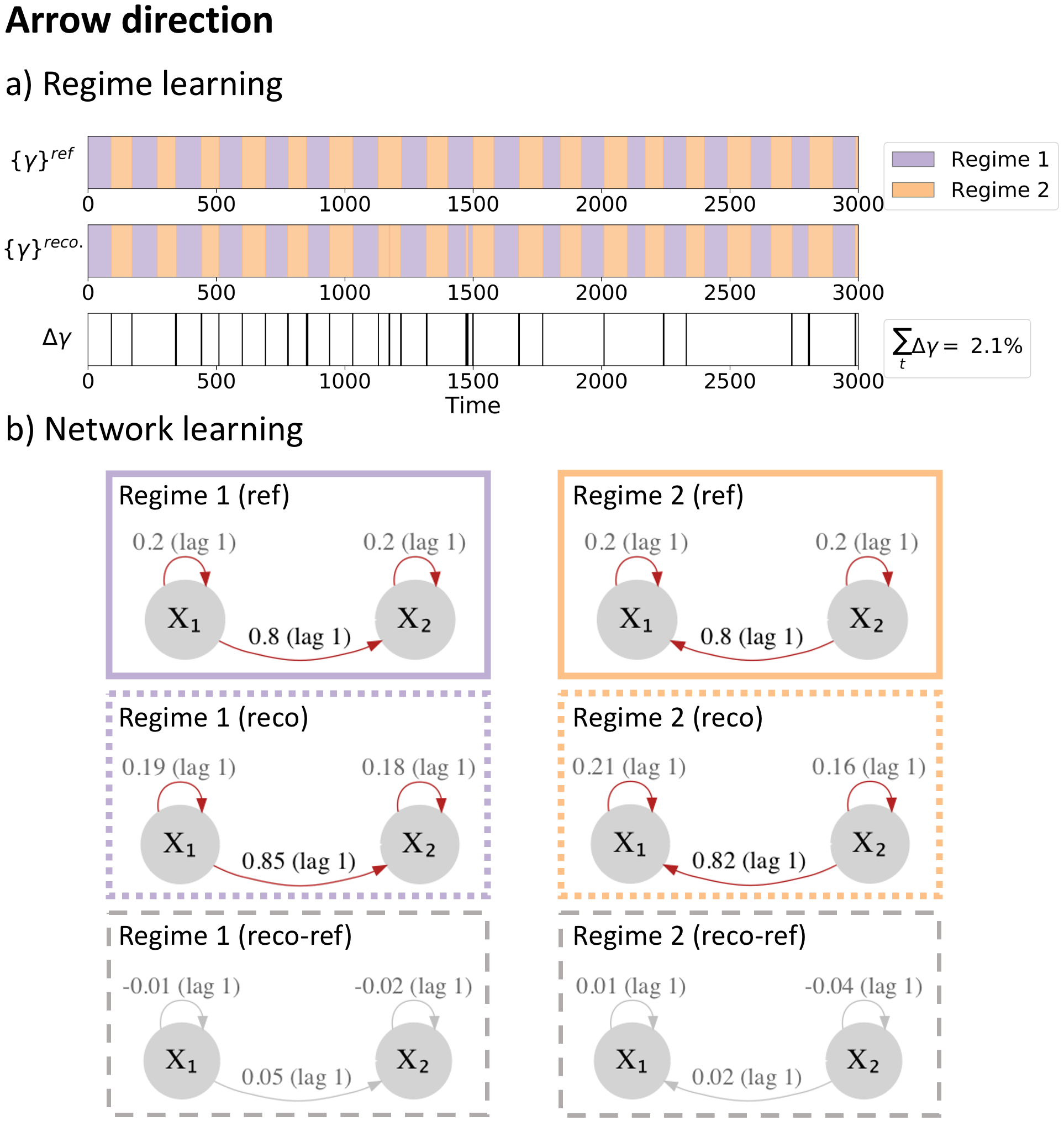}
\caption{Example case \textit{Arrow direction}. See description in Figure \ref{fig:signXY}. }
\label{fig:arrow}
\end{figure}

\begin{figure}
\center
\includegraphics[width=0.8\linewidth]{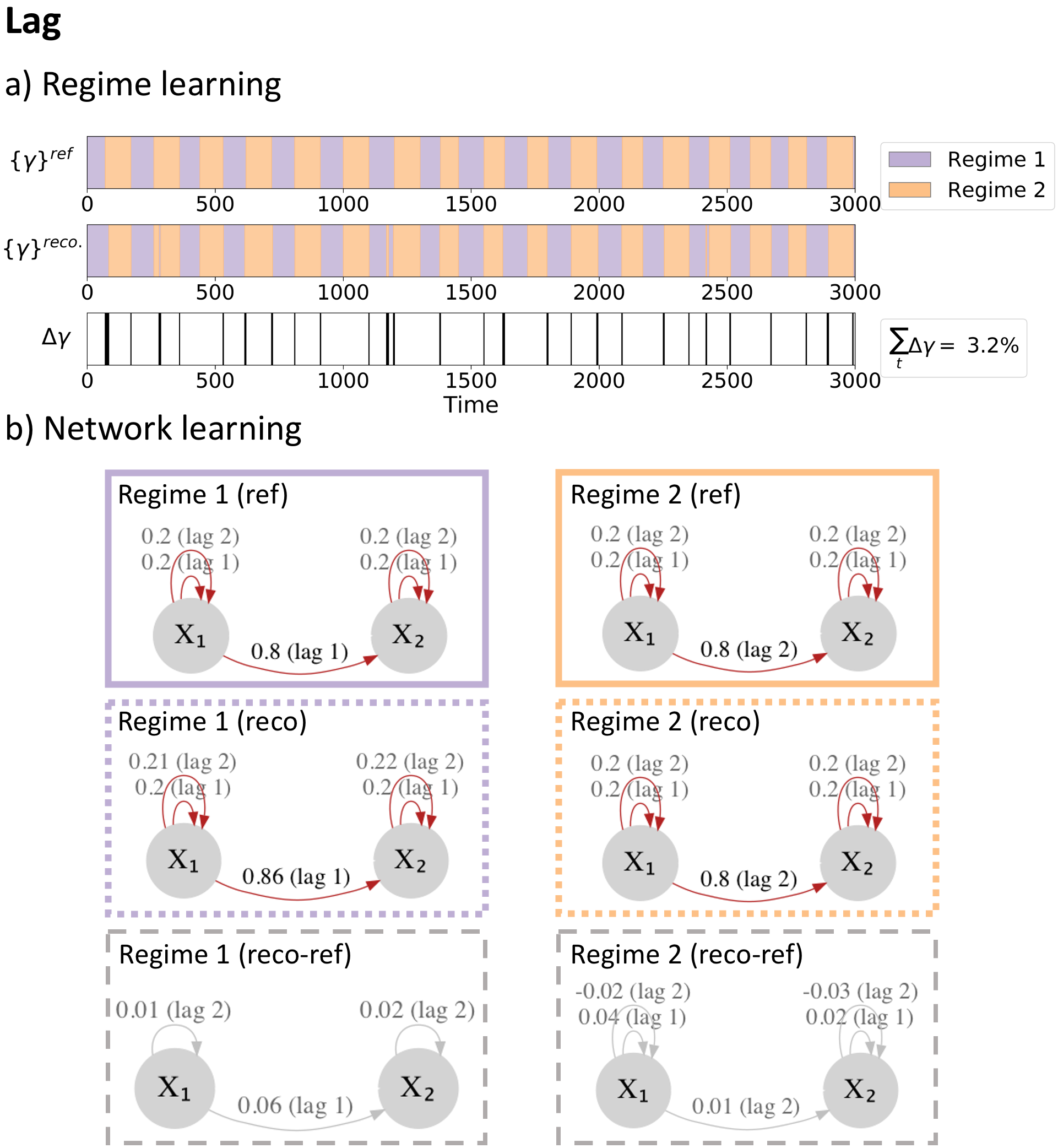}
\caption{Example case \textit{Lag}. See description in Figure \ref{fig:signXY}.}
\label{fig:lag}
\end{figure}

\begin{figure}
\center
\includegraphics[width=0.8\linewidth]{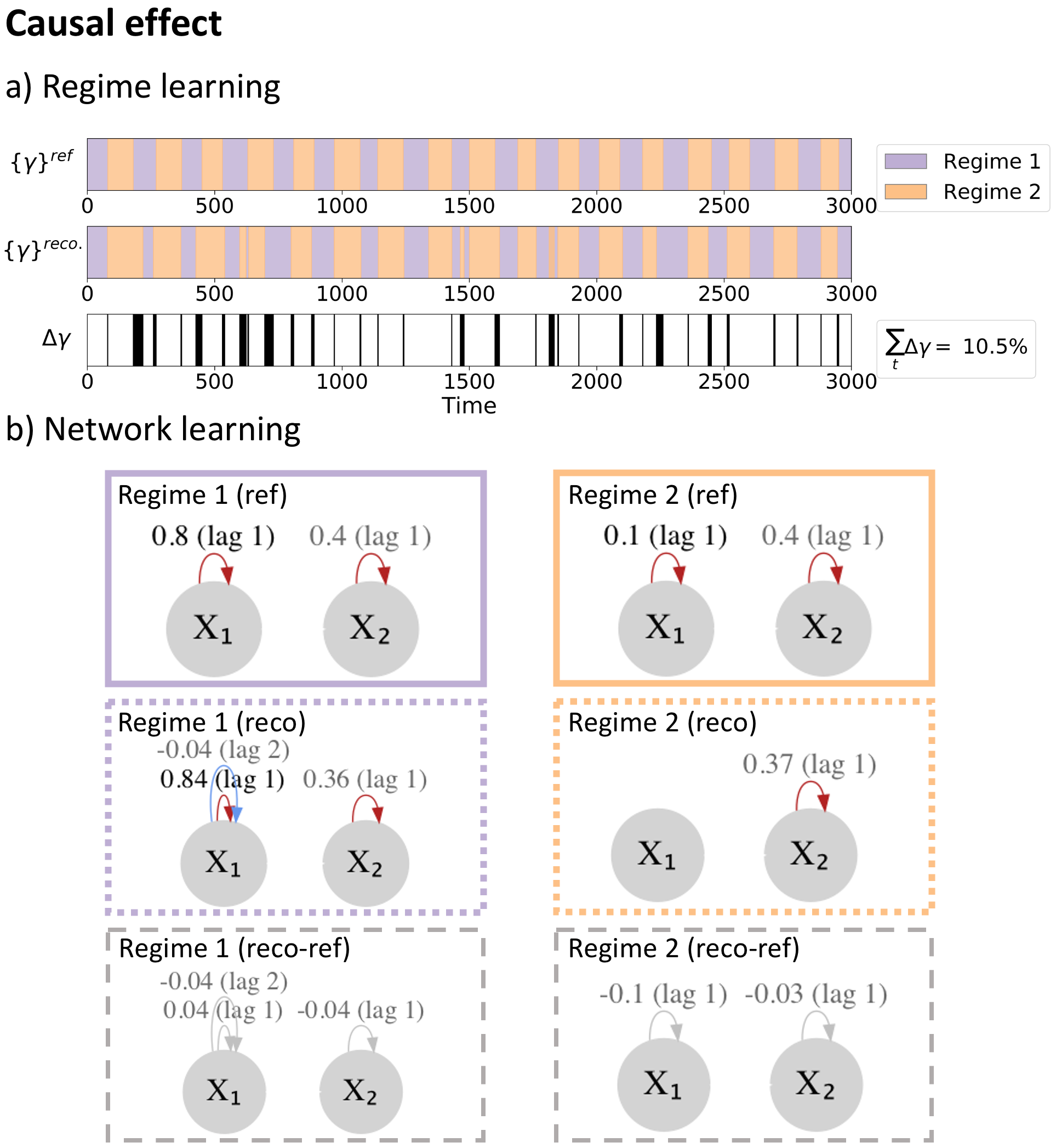}
\caption{Example case \textit{Causal effect}. See description in Figure \ref{fig:signXY}.}
\label{fig:causaleff}
\end{figure}

The ability of the proposed method to recover the networks and the regimes on the basis of the artificially designed time series are presented in the following. Figures \ref{fig:signXY}-\ref{fig:causaleff} present results for each case in Table~\ref{tab:linearexp}, focusing on one of the $N_R$ synthetic datasets. Table \ref{tab:ex_stats} shows summary statistics over all $N_R$ runs. 

The case \textit{sign $X^1 X^2$} is discussed in detail. The ground-truth regime evolution and networks are shown in the top part of panels \textit{a} and \textit{b} in Figure \ref{fig:signXY}; in the middle part of both panels their Regime-PCMCI reconstruction is shown; in the bottom part the difference between reconstructed and true regimes are presented to visually inspect the accuracy. The reconstructed regime assigning process for each regime matches the truth in 99.6\% of time steps (97\% average value over $N_R$, see Table \ref{tab:ex_stats}). The corresponding networks have all and only the correct links (TPR = $0.99$ and FPR = $0.01$ average value over $N_R$); their linear causal effect is also well estimated with each link correct up to $\pm 0.02$ ($N_R$-averaged error per link is $0.028$ (9\%)).

The other four cases are presented in Figures \ref{fig:signX}-\ref{fig:causaleff}. The case \textit{causal effect}, and to a lesser extent  \textit{lag change}, are hardest to detect. This is because the difference between the individual regimes and a mixed state of the two is not very large and thus the detection is more challenging. This adds to the general challenge of non-convexity of the functional we are optimising, which we mitigate by the annealing steps as mentioned in section \ref{sec:Modelselection}. A similar challenge is found for some high dimensional runs for which we refer to section \ref{sec:Highdim}. 

Table \ref{tab:ex_stats} shows the summary results over the $N_R$ realisations. The estimation errors are presented in terms of the regime assigning process (second column), the network structure (third to sixth column), the causal effects of links (seventh to tenth columns) and the overall reconstructed time series (last column). The second column, $\Delta \gamma \%$, is the average percentage of wrongly estimated time steps per regime (the lower the better, note that this value is the same for $k=1,2$, by construction).
In terms of networks, the link detection performance is evaluated via the true positive (TPR) and false positive rates (FPR). Further, we compare these with the reference FPR and TPR (superscript \textit{ref}) if PCMCI is run with the ground-truth regime variable known (but causal structure unknown). The accuracy in links' causal effects is assessed via ${\Delta \Phi}$, the average difference between the reconstructed linear coefficient and the reference values of the ground truth links. ${\Delta \Phi}$ is also expressed as percentage, i.e. each difference is weighted by the absolute value of the ground truth coefficient. The last column, $\hat \varepsilon$, is the expected prediction error per variable and per time step and is computed as $\hat \varepsilon = \sqrt{\mathbf{L}/(N_X T)}$ with $\mathbf{L}$ defined in Eq \eqref{eq:costfunctional_reg} and with $N_X$ and $T$ referring to the number of variables, here two, and the length of the time series respectively.
The precise definition of all the above statistics can be found in Appendix \ref{ap:stat_def}.

\begin{table*}[!ht]\center
\setlength{\tabcolsep}{8pt} % Default value: 6pt
\renewcommand{\arraystretch}{1.5} % Default value:1 
\begin{tabular}{ l | c| c  c| c c| c   c | c  c | c}% r }
%\hline
     \textbf{Example}& $\Delta\gamma \%$& $\text{TPR}_{\text{all}}$&$\text{TPR}^{\text{ref}}_{\text{all}}$& $\text{FPR}_{\text{all}}$&$\text{FPR}^{\text{ref}}_{\text{all}}$ &
     $\Delta \Phi$& $\Delta \Phi^{\text{ref}}$ &
     $\Delta \Phi$ $\%$& $\Delta \Phi^{\text{ref}}$ $\%$ &
       
     $\hat \epsilon$\\ %& $\%$ $\textbf{isminimal}$\\ 
\hline
  \textit{arrow direction}  & $3.0$   & $1.0$     &$1.0$   & $0.02$ & $0.01$ & $0.021$& $0.020$& $7.0$&$7.0$& $0.76$\\% &$92$ \\
    \textit{causal effect}  & $43.0$   & $0.81$     &$0.98$   & $0.11$ & $0.01$  & $0.286$& $0.020$& $120.0$&$10.0$& $0.68$\\% &$16$ \\
      \textit{lag } & $6.0$   & $0.98$     &$1.0$   & $0.04$ & $0.01$  & $0.027$& $0.018$& $11.0$&$8.0$& $0.68$\\% &$60$ \\
       \textit{sign $X^1$}  & $4.0$   & $0.98$     &$1.0$   & $0.03$ & $0.01$  & $0.033$& $0.016$& $10.0$&$6.0$& $0.65$\\% &$52$ \\
          \textit{sign $X^1X^2$}  & $3.0$   & $0.99$     &$1.0$   & $0.01$ & $0.01$  & $0.028$& $0.019$& $9.0$&$7.0$& $0.75$\\% &$70$ \\
\hline
  \end{tabular}
  \caption{Results for $N_K=2$ experiments averaged over $N_R=100$ realisations generated for each example described in Table \ref{tab:linearexp}.}
   \label{tab:ex_stats}
\end{table*}

In summary, Table \ref{tab:ex_stats} shows that :
\begin{itemize}
\item $\Delta \gamma \%$:  on average, the regime assigning process is reconstructed correctly in $ \sim 94\%$ of the time steps for all cases except \textit{causal effect}. The \textit{causal effect} and \textit{lag} examples are the hardest to infer, with causal effect being particularly deficient. In these examples a mixed-regime state (e.g., arising from assigning a considerable fraction of wrong time steps to a regime) is still quite close to any of the true   regimes. Therefore the algorithm struggles to decide which time steps belong to which regime, since they could fit both to some degree. Yet, there are 7 instances where $\Delta \gamma < 15 \%$ (one presented in Figure \ref{fig:causaleff}) and those, as expected from PCMCI, give very good network fit. We notice that these runs do not correspond to the lowest objective values of the $N_R$ set (i.e. better fit) which shows that runs that end up in mixed states can still fit the data quite well. Also, we notice that the causal effect setup reaches local minima in 16\% of the 100 runs, thus in 84\% of the runs the algorithm cannot easily find a stable solution which points at a weaker confidence in the output.

\item TPR: despite some errors in reconstructing the regime assigning process, the TPR is always very close to 1. This can indicate that the true signals, dynamic wise, are strong enough to be detectable.

\item FPR: Ideally the false positive rate should be upper-bounded by $\alpha=0.01$. This is also the case if we assume the correct regimes (see column $\text{FPR}^{\text{ref}}$). However, if the regimes are learned, in most of the examples the FPR value is higher due to errors in learning the regimes. If a wrong regime is learned, then both false positives and false negatives can occur. False negatives, i.e., missing links in the PC$_1$ step of PCMCI can lead to false positives in the MCI step.
\item $\Delta \Phi \%$: Errors in parents' detection (either due to false positives (FPR) or to false negatives (missed links, FNR = 1-TPR)) surely impact the estimation of link effects. Since the TPR and FPR are good, except for the causal effects case, we expect to obtain also good results for the linear coefficients.
This is indeed the case, as the difference  is of order $10^{-2}$ implying a relative error of about $10 \%$. 

\end{itemize}

\subsection{Low dimensional data with three underlying regimes}\label{sec:K3}
To illustrate how Regime-PCMCI deals with more than two regimes, we also considered a toy time series  based on three different causal regimes. It is of course possible to consider the case $N_K>3$, yet in  applications it is often desirable to infer a few prominent and relevant regimes rather than having too many that are not interpretable anymore. In other words, the aim is to avoid overfitting and to increase the information gain by reducing the complexity of the assumed  model (parsimony).
% table entries match examples

\subsubsection{Experiment settings}
The artificial time series is generated via a regime dependent causal graph that is designed by combining two of the regimes settings presented in section \ref{sec:Lowdim_linear}, namely \textit{sign $X^1X^2$} change and \textit{arrow inversion} (for details see Table \ref{tab:K3exp}). The regime assigning reference process $\{\Gamma\}^{\text{ref}}$ is generated by randomly choosing between different persistence lengths of $60$, $70$ and $80$ time-steps and iterating for $20$ times. The algorithm is run with free parameters in Table \ref{tab:linearlowdimruns_K3_settings}. 
 
\begin{table*}[htp!]
\small
\center
\setlength{\tabcolsep}{8pt} % Default value: 6pt
\renewcommand{\arraystretch}{1.5} % Default value:1 
\begin{tabular}{ l  |c | c | c |c|c| c }
    \textbf{example} & $k=1$ &$k=2$ &$k=3$ & ${\Phi}^j_1(i,\tau)^{\text{ref}}$ & $\{{\Phi}^j_2(i,\tau)\}^{\text{ref}}$ &$\{{\Phi}^j_3(i,\tau)\}^{\text{ref}}$\\
    \hline
    \textit{sign $X^1X^2$} &$X^1 \xrightarrow{|a|} X^2$ & $X^1 \xrightarrow{-|a|} X^2$& $X^ 2\xrightarrow{|a|} X^1$& $\{{\Phi}^2_1(1,1)\}^{\text{ref}}=0.8$ & $\{{\Phi}^2_2(1,1)\}^{\text{ref}}=-0.8$ & $\{{\Phi}^1_3(2,1)\}^{\text{ref}}=0.8$ \\
   and \textit{arrow}  & & && $\{{\Phi}^1_1(1,1)\}^{\text{ref}}=0.2$&  $\{{\Phi}^1_2(1,1)\}^{\text{ref}}=0.2$&$\{{\Phi}^1_3(1,1)\}^{\text{ref}}=0.2$  \\
     \textit{direction}   & & && $\{{\Phi}^2_1(2,1)\}^{\text{ref}}=0.2$&  $\{{\Phi}^2_2(2,1)\}^{\text{ref}}=0.2$& $\{{\Phi}^2_3(2,1)\}^{\text{ref}}=0.2$ \\
        \hline
  \end{tabular}
  \caption{Artificial model configuration for a low dimensional example with $N_K=3$ underlying regimes.}
   \label{tab:K3exp}
\end{table*}

\begin{table}
\center
\setlength{\tabcolsep}{8pt} % Default value: 6pt
\renewcommand{\arraystretch}{1.5} % Default value:1 
\begin{tabular}{ c  c  c c | c  c c c}
%\hline
  CI test & $\tau_{\max}$ &  $\alpha$ & $\alpha_{\rm PC}$ & $N_K$ & $N_C$ & $N_Q$ & $N_A$\\ 
\hline
%\hline
    ParCorr  & $3$   & $0.01$     &$0.2$     & $3$        & $40$ & $20$  & $50$   \\
%\hline
  \end{tabular}
  \caption{Method parameters for a low dimensional example with $N_K=3$ underlying regimes.}
   \label{tab:linearlowdimruns_K3_settings}
\end{table}

\subsubsection{Results}
\begin{figure}
\center
\includegraphics[width=0.8\linewidth]{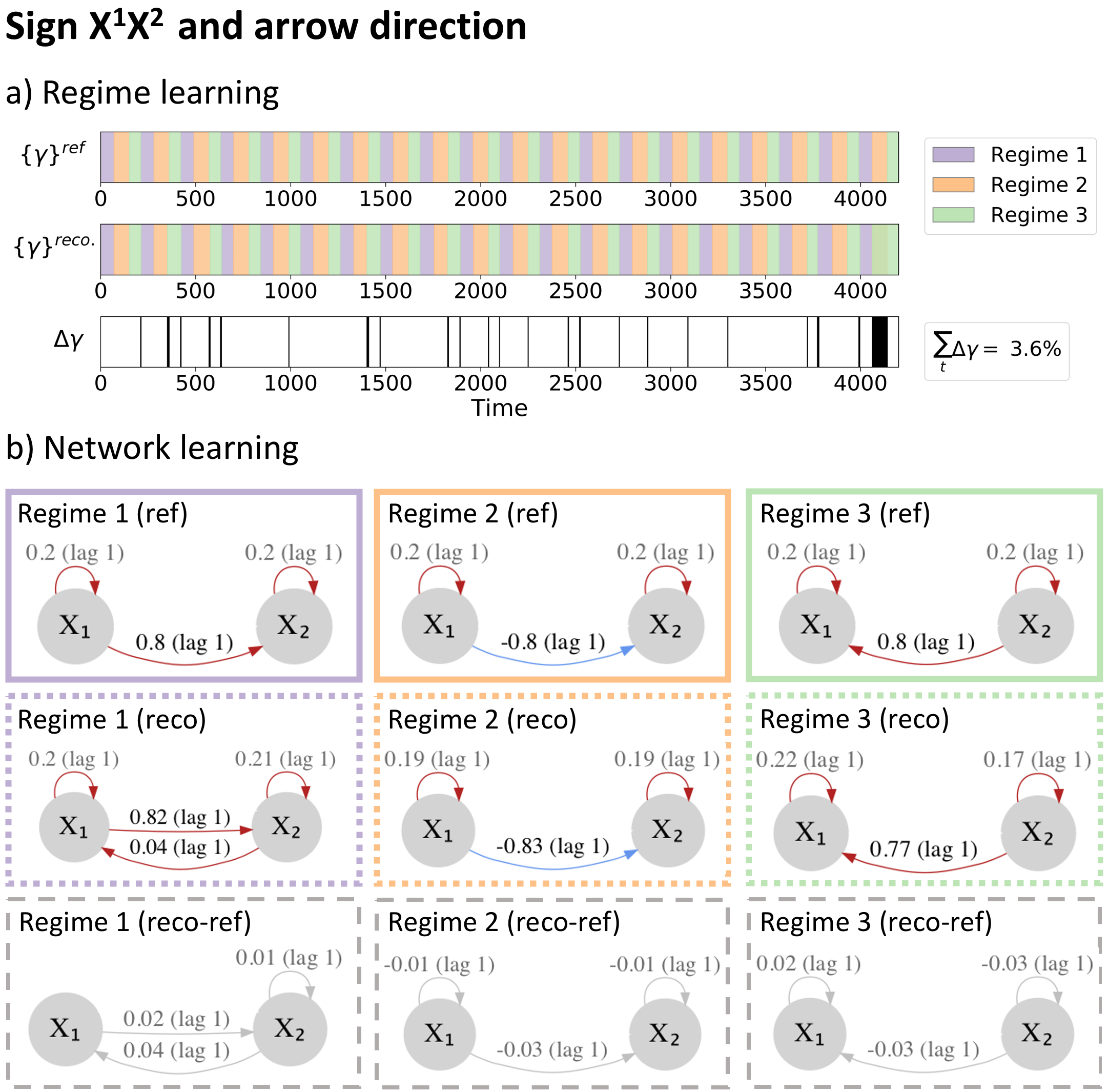}
\caption{Example with $N_k=3$ regimes for the case \textit{Sign $X^1X^2$ and arrow direction}. See description in Figure \ref{fig:signXY} but with three regimes. }
\label{fig:NK3}
\end{figure}

Figure \ref{fig:NK3} shows the results. 
% obtained regime assignment and coefficient $\phi_k$ from Regime-PCMCI compared to the values used to generate the data. 
There are only minimal deviations from the true reference values which confirms that the proposed method is capable to deal with $N_K>2$. This also holds for the summary results over $N_R=100$ runs presented in Table \ref{tab:K3_ex_stats}. Yet, it is important to note that we chose a combination of causal graphs that performed well for $N_K=2$, i.e, causal effect changes would also be difficult to detect for $N_K=3$.

\begin{table*}[!ht]\center
\setlength{\tabcolsep}{8pt} % Default value: 6pt
\renewcommand{\arraystretch}{1.5} % Default value:1 
\begin{tabular}{ l | c  c| c c| c   c | c  c | c}% r }
%\hline
      $\Delta\gamma \%$& $\text{TPR}_{\text{all}}$&$\text{TPR}^{\text{ref}}_{\text{all}}$& $\text{FPR}_{\text{all}}$&$\text{FPR}^{\text{ref}}_{\text{all}}$ &
     $\Delta \Phi$& $\Delta \Phi^{\text{ref}}$ &
     $\Delta \Phi$ $\%$& $\Delta \Phi^{\text{ref}}$ $\%$ &
       
     $\hat \epsilon$\\ %& $\%$ $\textbf{isminimal}$\\ 
\hline
   $4.0$   & $0.98$     &$1.0$   & $0.05$ & $0.01$ & $0.033$& $0.020$& $10.0$&$7.0$& $0.5$\\
\hline
  \end{tabular}
  
  \caption{Results for $N_K=3$ experiments averaged over $N_R=100$ realisations generated for each example described in Table \ref{tab:K3exp}.}
   \label{tab:K3_ex_stats}
\end{table*}

\subsection{Regime parameter selection}
\label{sec:K_selection}
\begin{figure}
\center
{\includegraphics[width=.8\linewidth]{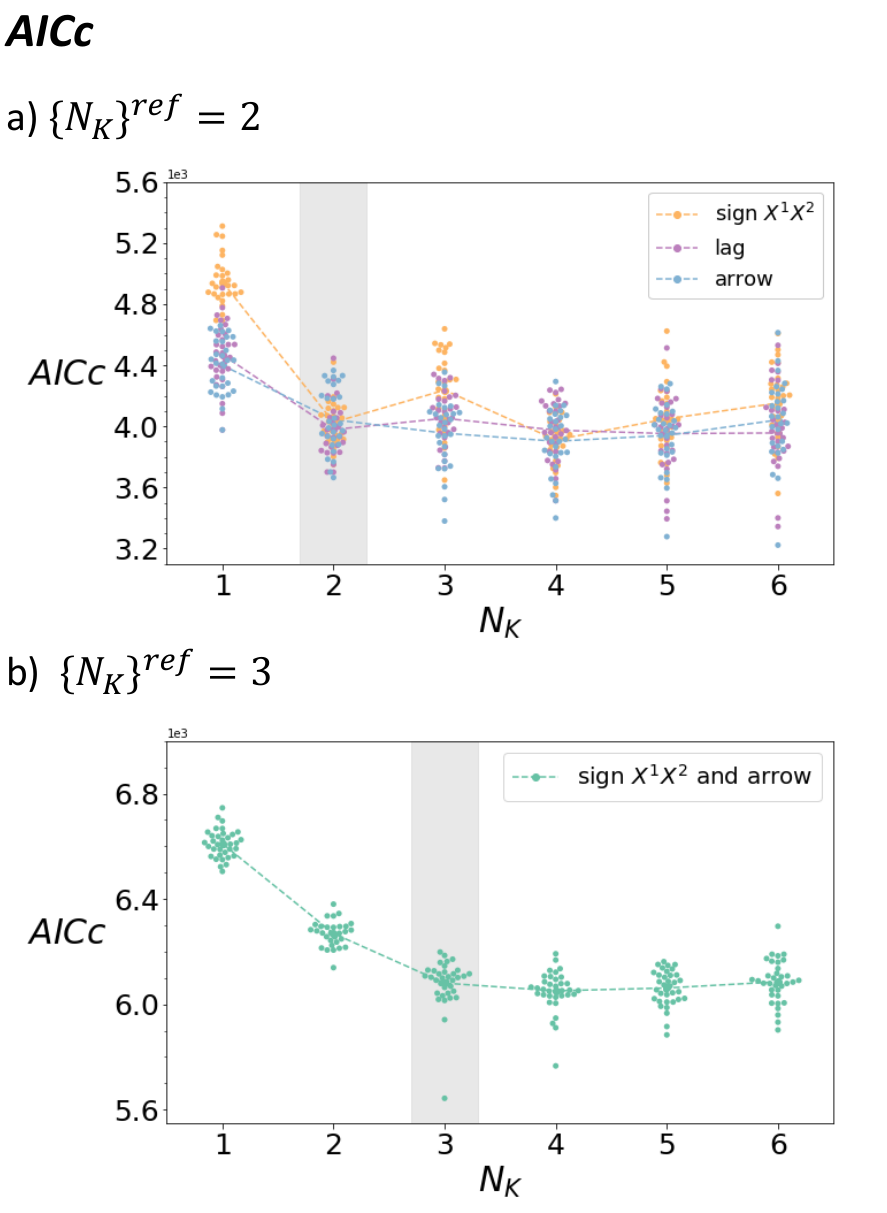}}
\caption{Numerical investigation of AICc values for runs with different $N_K$ and (a) $\{N_K\}^{\text{ref}}=2$ for three networks examples (\textit{sign $X^1X^2$, arrow} and \textit{lag} change) and (b) $\{N_K\}^{\text{ref}}=3$ for the \textit{sign $X^1X^2$ and arrow} change example. In each example, individual dots represent the value attained by the $N_R=29$ runs, and the dashed line goes through the mean values of each set. The vertical grey bar highlights the ground-truth number of regimes $\{N_K\}^\text{ref}$. }
\label{fig:AICc}
\end{figure}

We investigate how parameter selection of the number of regimes affects the results by means of the AICc scores defined in (\ref{eq:AICc}). We investigate two test scenarios of $\{N_K\}^{\text{ref}}=2,3$ for a selection of the examples defined in the previous sections \ref{sec:Lowdim_linear} and \ref{sec:K3}. The PCMCI parameters are as in those sections while $N_R=29$, $N_Q=20$ and $N_A=20$. The resulting AICc values are displayed in Figure \ref{fig:AICc}. The $N_C$ value is changed adaptively for each $N_K$ to ensure a similar $N_M$ value for the different number of regimes, i.e., 
\begin{equation}
N_C(N_K)=\{N_C^{\text{ref}}\}\{N_K\}^{\text{ref}}/N_K
\end{equation}
for $N_K>\{N_K\}^{\text{ref}}$. The reference value for the number of switches is on average (due to randomisation of $\{\Gamma\}^{\text{ref}}$) $\{N_C^{\text{ref}}\}=40$ for both $\{N_K\}^{\text{ref}}=2,3$.

We note that the lowest $N_K$ at which the AICc plateaus is the ground-truth one. The plateau itself occurs due to the fact that only the links with non-zero causal effect values are counted towards the number of parameters. Thus a higher number of regimes $N_K$ does not necessarily result in an increase of the total number of parameters. In other words the penalisation is not becoming stronger with higher values of $N_K$. Concluding, it is clearly visible that no significant improvement is gained by increasing the number of $N_K$ beyond the reference number of regimes. Since the entry point to the plateau reveals the reference number of regimes, it seems possible to face scenarios where the true number of regimes is unknown.

%%%%%%%%%%%%%%%%%%%%%%%%%%%%%%%%%%%%%%%%%%%%%%%%%%%%%%
%%%              High dimensional               %%%%%%
%%%%%%%%%%%%%%%%%%%%%%%%%%%%%%%%%%%%%%%%%%%%%%%%%%%%%%%

\subsection{High dimensional linear network}
\label{sec:Highdim}
In this section the algorithm is evaluated on high-dimensional datasets, with each dataset consisting of $N_X=10$ interacting variables. The background regimes are generated with two regular alternating regimes of 300 time steps each, for a total length $T = 15,000$. The network structures are randomly generated from a family of linear networks defined via the parameters shown in Table \ref{tab:highdim_netparam}, where $L$ is the number of randomly drawn cross variable links with random coefficients from the third column. Note that each variable is also auto-linked at lag 1 with coefficient randomly drawn from the fourth column. The time series $\mathbf{x}_t \in \mathrm{R}^{10}$ are generated with model~\eqref{eq:toy_timeseries} and for $N_R=70$ realisations. Regime-PCMCI is then run with the settings shown in Table~\ref{tab:highdimruns_settings}.

\begin{table}[ht!]
\center
\setlength{\tabcolsep}{8pt} % Default value: 6pt
\renewcommand{\arraystretch}{1.5} % Default value:1 
\begin{tabular}{ccccc} 
$N_X$ & L & $\Phi^j_k(i,\tau)$ & $\Phi^i_k(i,\tau)$& max lag   \\ 
\hline 
10 & 30 & [-0.4, 0.4]& [ 0.2, 0.5, 0.9] & 3 \\ 
\end{tabular} 
\caption{High dimensional networks parameters}
\label{tab:highdim_netparam}
\end{table}

\begin{table}[h!]\center
\setlength{\tabcolsep}{8pt} % Default value: 6pt
\renewcommand{\arraystretch}{1.5} % Default value:1 

\begin{tabular}{ c  c  c c | c  c c c}
%\hline
  CI test & $\tau_{\max}$ &  $\alpha$ & $\alpha_{\rm PC}$ & $N_K$ & $N_C$ & $N_Q$ & $N_A$\\ 
\hline
%\hline
    ParCorr  & $4$   & $0.05$     &$0.2$     & $2$        & $49$ & $30$  & $50$   \\
%\hline
  \end{tabular}

  \caption{Method parameters for high dimensional experiments with two underlying regimes.}
   \label{tab:highdimruns_settings}
\end{table}

The results are shown in Table \ref{tab:HD_ex_stats}, which is structured like Table \ref{tab:ex_stats}. Regime-PCMCI performs very well even in this challenging setting. 
Notably, individual runs can perform extremely well, with $\Delta \gamma $ reaching as low as $0.02\%$, and a total of 53 runs below total average of $\Delta \gamma =11.7 \%$ (second row in Table \ref{tab:HD_ex_stats}). The other 7 runs are responsible for most of the deviation of the average statistics from the reference values (first row).

As in the \textit{causal effect} case, there is a mismatch between runs with the lowest prediction errors $\hat \varepsilon$ and the lowest error on regime-assigning process $\Delta \gamma$, meaning that we cannot use a filtering on $\hat \varepsilon$ to find the best performing runs. This behaviour can be explained from the tendency of the algorithm to still over-fit when too many degrees of freedom are available, as well as from the complexity of distinguishing different causal effects (a challenge already manifested in the \textit{causal effect} case).

\begin{table*}[!ht]\center
\setlength{\tabcolsep}{8pt} % Default value: 6pt
\renewcommand{\arraystretch}{1.5} % Default value:1 
\begin{tabular}{ l | c| c  c| c c| c   c | c  c | c | r }
%\hline
    Selection & $\Delta\gamma \%$& $\text{TPR}_{\text{cros}}$&$\text{TPR}^{\text{ref}}_{\text{cros}}$& $\text{FPR}_{\text{cros}}$&$\text{FPR}^{\text{ref}}_{\text{cros}}$ &
     $\Delta \Phi$& $\Delta \Phi^{\text{ref}}$ &
     $\Delta \Phi$ $\%$& $\Delta \Phi^{\text{ref}}$ $\%$ & $\hat \epsilon$ &n. runs \\
\hline
  all&                        $11.7$  & $0.94$ &$1.0$   & $0.18$ & $0.08$ & $0.059$& $0.005$& $16.0$&$1.5$& 0.85 & 70 \\
  $\Delta \gamma < 11.7$ \% & $0.19$ & $1.0$ & $1.0$ & $0.08$ & $0.07$ & $0.006$ & $0.005$ & $1.8$ & $1.5$  &0.70 &  $53$ \\
\hline
  \end{tabular}
  
  \caption{Results for high-dimensional experiments over $N_R=70$ realisations generated for each example described in Table \ref{tab:highdim_netparam}.}
   \label{tab:HD_ex_stats}
\end{table*}

\subsection{Computational complexity}
Table \ref{tab:runstats} shows some indicators of performance of the method: the fraction of $N_R$ runs that correspond to a (local) minima, %the percentage of annealing per each run that reach a minima and 
the average number of q-iterations needed to reach a local minima and the runtime for the whole $N_R$ set of runs  % and The mean value of the prediction error across all $N_R$.
(the code run parallel over the $N_A$ annealings and using 4 to 6 CPUs per job). \\
Most of the examples reach local minima in more than 50\% of the $N_R$ runs, while the percentage is very low for \textit{causal effect} (second column). We note that examples with high percentage of local minima correspond also to quick convergence in terms of iterations steps (third column). They are also associated with better regimes reconstruction (see Table \ref{tab:ex_stats}, \ref{tab:K3_ex_stats},\ref{tab:HD_ex_stats}), confirming that a clear cost functional minimum (as shown from the second and third column) is linked to better detection. Finally, the runtime is quite fast: the low dimensional examples take between $10$ and $20$ minutes for $N_K=2$ and 45 minutes for $N_K=3$ to complete $100$ runs. The high dimensional example takes just below $3$ hours for $70$ runs.

\begin{table*}[!ht]\center
\setlength{\tabcolsep}{8pt} % Default value: 6pt
\renewcommand{\arraystretch}{1.5} % Default value:1 
\begin{tabular}{ l |  c  c  c  }
%\hline
    \textbf{Example} & n. local minima/$N_R$ \%  & iterations to minima & runtime ($s$) \\ %& n. local minima per run / $N_A$ \%
\hline
\textit{arrow direction}  &  $92$ \%  & $7$ & 600  \\%&  $98$ \% 
\textit{causal effect}  & $16$ \% & $13$ & 970 \\ %& $32$ \%& 
\textit{lag}  & $60$ \% & $11$ & 1,130\\%$84$ \%& 
\textit{sign} $X^1$ & $52$\% & $12$ & 970 \\%& $74$ \%
\textit{sign} $X^1X^2$  & $70$ \% & $9$ & 700 \\%& $93$ \%
\textit{sign} $X^1X^2$ and \textit{arrow} & $56$\% & $10$ & 2,670 \\%&$80$ \% 
\textit{high dimensional} & $92\%$  & $6$ & 10,780\\%& $97$ \%
\hline
  \end{tabular}
  
  \caption{Summary performance statistics of all examples. The third column is the average value over the respective $N_R$. }
   \label{tab:runstats}
\end{table*}

\section{A real-world example: the effect of \\ El Ni\~no Southern Oscillation on Indian rainfall}\label{sec:Climatedataex}

We finally test the performance of Regime-PCMCI on real-world data, and apply it to address the non-stationary relationship of  El Ni\~no Southern Oscillation (ENSO) and all-India rainfall (AIR) mentioned in the introduction. We are interested in if, for given time series of ENSO and AIR, our method is able to distinguish between the winter and summer months, i.e. the background-regimes, and to detect a reported link from ENSO to AIR during summer. 

This example can be considered a difficult case since the expected signal from ENSO to AIR is likely small compared to natural variability \cite{WebsterPalmer1997}. Further, climate data is typically very noisy with causal relationships being diluted by other, often unknown processes given a complex coupled climate system \cite{Williamstal2017}.  

Our input data consist of monthly observations of ENSO  and AIR, for the years 1871 to 2016, resulting in two time series consisting of 1740 monthly values each. More precisely, ENSO is represented by the so-called relative Nino3.4 index  provided by the National Oceanic and Atmospheric Administration (NOAA) \cite{Huang2017}\footnote{ \url{http://climexp.climexp-knmi.surf-hosted.nl/getindices.cgi?WMO=NCDCData/ersst_nino3.4a_rel&STATION=NINO3.4_rel&TYPE=i&id=someone@somewhere}}. %It is based on reconstructed sea surface temperatures (SSTs) over the tropical Pacific, .... anomalies normalized to 1981-2010 and with the 20S-20N average SST subtracted, i.e. global warming is removed.\mk{is it needed o describe index in detail?}. 
Data for AIR anomalies (with the climatology subtracted) are provided by the Indian Institute of Tropical Meteorology (IITM) \cite{AIRdata}\footnote{  \url{http://climexp.climexp-knmi.surf-hosted.nl/getindices.cgi?WMO=IITMData/ALLIN&STATION=All-India_Rainfall&TYPE=p&id=someone@somewhere}}.
We choose the following parameters of Regime-PCMCI: For the regime part, we set $N_K = 2$ and $N_C = 292$, which is equivalent to assuming two seasons per year. For the PCMCI settings, we use a significance level $\alpha=0.01$ ($\alpha_{\rm PC}=0.2$). Further, we use a maximum time-lag of two months, i.e. $\tau_{\max}= 2$. The optimisation is run $N_A=100$ annealing times, to span many local minima, with each annealing allowed for up to $N_Q=100$ iteration steps to converge. 

\begin{figure}
\center
{\includegraphics[width=0.8\linewidth]{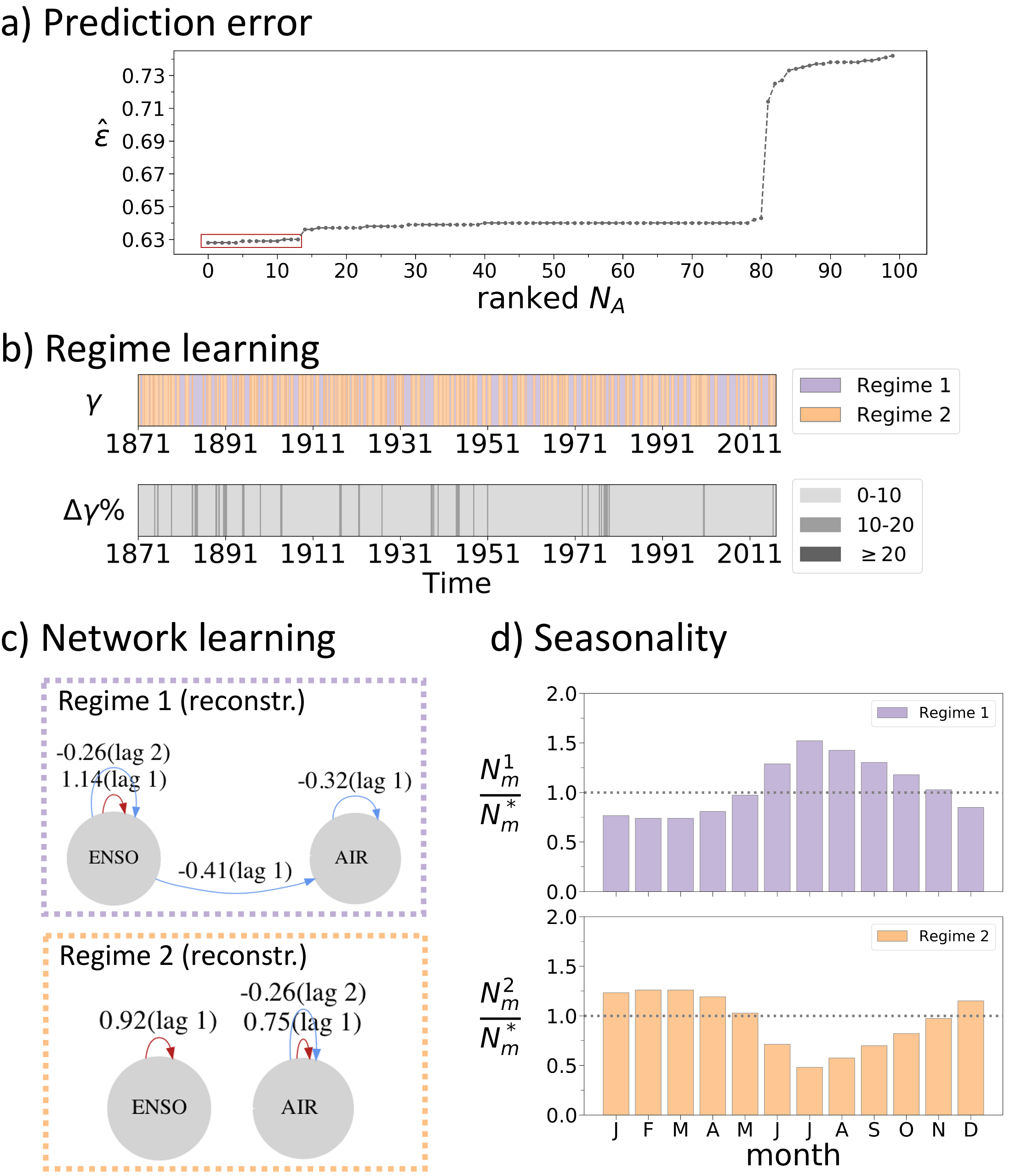}}
\caption{Climate example. (a) Prediction error for each annealing step in ascending order, lowest 13 annealings highlighted in red box. All the other panels refer to this selection. (b) Regime learning: regime-assigning process corresponding to the best annealing (rank 0) (top) and departure from this estimate of the remaining best 12 annealings (in percentage difference). (c) Network learning: mean networks per regime, each causal effect is the mean of of the corresponding coefficient in the individual 13 annealings. (d) Seasonality of the regimes: Number of years per month $m$ assigned to each regime ($N^{k}_m$), normalised by $N^*_m$, which refers to the expected number of months assigned to a given regime if one assumes equal probability $1/N_K$ of assigning a month to one of the two regimes. Thus, here, $N^*_m = 13 \cdot T /(12 * N_K)$).}
\label{fig:climate_ex}
\end{figure}

Among the annealing steps, which correspond to different random initial guesses on the regime-assigning process $\Gamma$, some clearly performed better in terms of fitting the data. We estimate the average prediction error associated with each annealing, $\hat \varepsilon$ (\ref{eq:predition_err}), and Figure \ref{fig:climate_ex},a shows it for all annealings (ranked according to $\hat \varepsilon$). A red box highlights the top performing cluster (13 runs). 

All of the top 13 annealing find a link from ENSO to AIR during one of their two regimes only (for simplicity hereafter called regime 1). In the following we present results averaged over these annealings and plot links that surpass a strength of $0.1$.

The causal link from ENSO to AIR in regime 1 has an average standardized linear effect of $-0.4$, meaning that a one standard deviation increase in ENSO results in a reduction of $0.4$ standard deviations in AIR (Figure \ref{fig:climate_ex},c) . This negative dependence is well documented in the literature \cite{WebsterPalmer1997}. During regime 2, in contrast, ENSO and AIR are, on average, almost independent, with only a very weak link ($- 0.05$, not shown) detected from AIR to ENSO. 
More importantly, our results indicate a clear seasonal dependence. Figure \ref{fig:climate_ex},d shows the number of months assigned to each regime (normalised by the number one would expect on the hypothesis of no seasonality, see figure caption). A clear peak in summer months is found for regime 1. More precisely, most of the months between June to September are assigned to regime 1 ($70\%$). These are the months in which the Indian summer Monsoon is active and for which a robust influence from ENSO has been shown.  In contrast, months assigned to regime 2 are predominantly winter months ($60\%$ of all December to March months). Thus, despite the relatively weak mean causal effect of ENSO on AIR during summer, and the large inter-annual variability, our algorithm successfully reconstructed this well-documented relationship given all-year time series of ENSO and AIR. 

Overall, these results are promising and show the potential of Regime-PCMCI to detect regime-dependent causal structures in a system as complex as the climate system. On the other hand, it also shows that domain knowledge is required to assure a suitable choice of parameters ($N_C$ and $N_K$) and an interpretation of the results. This is yet a common caveat to many data-driven approaches, which we nevertheless want to stress strongly.

\section{Discussion and conclusions}
Causal discovery is emerging as an important framework across many disciplines in science and engineering, but each discipline has particular challenges that novel  methods need to address \cite{Runge2014a}.
We introduced a novel method, Regime-PCMCI, to learn regime-dependent causal relations, overcoming one of the key drawbacks of current causal recovery methods. The performance of  Regime-PCMCI was analysed for many different artificially generated causal scenarios and for varying regimes showing that the method covers a wide range of settings (see Figures \ref{fig:signXY}-\ref{fig:lag} and Table \ref{tab:ex_stats}). The performance of the algorithm is maintained also for high-dimensional settings with 10 variables (see Table~\ref{tab:HD_ex_stats}) as well as for more than two regimes (see Figure~\ref{fig:NK3} and Table~\ref{tab:K3_ex_stats}). 
We found limitations of the method for the case where only the causal effect strength of a link changes between regimes (see Figure~\ref{fig:causaleff}) which seems to be hard to detect with our optimisation scheme and requires further investigation.
Further, the capability of  Regime-PCMCI was verified by means of a well documented climate example using real data of ENSO and Indian rainfall (see Figure \ref{fig:climate_ex}). Overall, the proposed method presents itself as a promising approach in the context of nonlinear causal links manifested in regime changes in time. 

Note that a causal interpretation of estimated links in our observational causal discovery framework still assumes causal sufficiency, that is, no unobserved common causes. However, estimated non-causality (zero coefficients) do not require this assumption \cite{jdw:Runge2018a} and can be interpreted as an absence of a causal relation already under the weaker Faithfulness assumption \cite{Runge2019b}. While for PCMCI asymptotic consistency was shown\cite{Runge2019b}, this is a more difficult task for Regime-PCMCI and deferred to further research.

% \subsection{Outlook}
There are several interesting aspects that could be explored in the future, building on the present work. These extensions can build on other causal discovery algorithms or extensions of PCMCI in the causal discovery step of our method. For example, as already discussed in section \ref{sec:nonlin}, the PCMCI algorithm allows for nonlinear causal links \cite{Runge2019b} and thus a nonlinear extension of the Regime-PCMCI is a logical next step. Recent extensions of PCMCI to the case of not only lagged, but also contemporaneous causal relations can also be integrated \cite{Runge2020a}.
Moreover, potentially it is also possible to better capture the causal effect-case and it might be possible to learn a regime-dependence of the noise term.

% Our approach could therefore be extended by combining latent causal discovery methods with our regime assignment procedure instead of PCMCI.

With respect to applications, it would be interesting to utilise the proposed method to study other links in the climate system that are likely regime-dependent, but less understood than the presented El Ni\~no-Indian rainfall example.

\section{Author's contribution}{E.S., J.dW., M.K. and J.R. designed the research, E.S. mainly performed the research, E.S., J.dW., M.K. and J.R. analyzed the results and wrote the manuscript.}

\section*{Acknowledgements}
E.S. was supported by the Centre for Doctoral Training in Mathematics of Planet Earth, UK EPSRC funded (grant EP/L016613/1). M.K. and J.dW. have been partially funded by the ERC Advanced Grant ACRCC (grant 339390). The research of J.dW. has been partially funded by Deutsche Forschungsgemeinschaft (DFG) (SFB1294/1-318763901) and by the Simons CRM Scholar-in-Residence Program. M.K. has received funding from the European Union’s Horizon 2020 research and innovation programme under the Marie Skłodowska-Curie grant agreement (No 841902). The optimisation software by Gurobi\texttrademark  was used for this work. The authors would like to thank Giorgia di Capua for discussions on the climate example. 

\section*{Data Availability Statement}
\noindent
The data that support the analysis of the first four sections of this study have been synthetically generated by the authors and can be fully reproduced using the equations and parameters described in the article. The data that support the findings of the last section of this study are openly available on the KNMI Climate Explorer at \url{https://climexp.knmi.nl/} (specific urls and original data source in references). Regime-PCMCI is part of the open-source Python package \texttt{tigramte} available at \url{https://github.com/jakobrunge/tigramite}.

\appendix

\section{Definition of result statistics}
\label{ap:stat_def}
\noindent
The definitions for the statistics presented in Tables \ref{tab:ex_stats}, \ref{tab:K3_ex_stats} and \ref{tab:HD_ex_stats} is outlined in the following.

\subsection{Regime assigning process}
\vspace{-15pt}
\begin{equation*}\label{eq:Dgamma_rel}
\Delta \gamma (\%)  = \frac{ \sum_{t=\tau_{\max}}^{T} | \{ \gamma_k(t) \}^{reco.}- \{ \gamma_k(t) \} ^{ref}|}{T-\tau_{max}}  \times 100 
\end{equation*}

\subsection{Link detection} 
%\subsubsection*{TPR}
\textbf{\textit{TPR}}
\begin{equation*}
\text{TPR} = \frac{{\text{TP}}_{\text{X}}} {{\text{P}}_{\text{X}}} 
\end{equation*}
Over the cross-variables links (in Tables \ref{tab:HD_ex_stats}): 

\begin{eqnarray}
\text{TP}_{\text{cros}}  = | \{ (i,j,\tau) : \{{\Phi}^j_k(i,\tau) \}^{reco.}\neq 0  \And&& \nonumber \\
\{ {\Phi}^j_k(i,\tau) \}^{ref}  \neq 0 \And i \neq j \} | && \nonumber 
\end{eqnarray}
\begin{eqnarray}
\text{P}_{\text{cros}} = { |\{ (i,j,\tau) : \{{\Phi}^j_k(i,\tau)\}^{ref}\neq 0  \And i\neq j \}|}&& \nonumber      
\end{eqnarray}

And over all links (in Tables \ref{tab:ex_stats} and \ref{tab:K3_ex_stats}):
\begin{eqnarray}
\text{TP}_{\text{all}}  = | \{ (i,j,\tau) : \{{\Phi}^j_k(i,\tau) \}^{reco.}\neq 0  \And &&  \nonumber\\
 \{ {\Phi}^j_k(i,\tau) \}^{ref}  \neq 0 \} |&& \nonumber  
 \end{eqnarray}
\begin{eqnarray}
\text{P}_{\text{all}} = { |\{ (i,j,\tau) : \{{\Phi}^j_k(i,\tau)\}^{ref}\neq 0 \}|}&& \nonumber    
\end{eqnarray}

%\subsubsection*{FPR}
\textbf{\textit{FPR}}
\begin{equation*}
\text{FPR} = \frac{{\text{FP}}_{\text{X}}} {{\text{N}}_{\text{X}}} 
\end{equation*}
Over the cross-variables links (in Tables \ref{tab:HD_ex_stats}): 
\begin{eqnarray}
\text{FP}_{\text{cros}}   = | \{ (i,j,\tau) : \{{\Phi}^j_k(i,\tau) \}^{reco.}\neq 0  \And &&      \nonumber    \\
\{ {\Phi}^j_k(i,\tau) \}^{ref}  = 0 \And i \neq j \} | &&      \nonumber    
\end{eqnarray}
\begin{eqnarray}
\text{N}_{\text{cros}}  = { |\{ (i,j,\tau) : \{{\Phi}^j_k(i,\tau)\}^{ref} = 0  \And i\neq j \}|}&&      \nonumber    
\end{eqnarray}

And over all links (in Tables \ref{tab:ex_stats} and \ref{tab:K3_ex_stats}):
\begin{eqnarray}
\text{FP}_{\text{all}} = | \{ (i,j,\tau) : \{{\Phi}^j_k(i,\tau) \}^{reco.}\neq 0  \And &&      \nonumber \\
 \{ {\Phi}^j_k(i,\tau) \}^{ref}  = 0 \} |&&      \nonumber   
 \end{eqnarray}
\begin{eqnarray}
\text{N}_{\text{all}} = { |\{ (i,j,\tau) : \{{\Phi}^j_k(i,\tau)\}^{ref}= 0 \}|}&&      \nonumber    
\end{eqnarray}

\subsection{Link coefficients}
\vspace{-15pt}
\begin{equation*}%\label{eq:DPhi_groundtruth}
 {\Delta \Phi} = \frac{1}{N_K} \sum_{k=1}^{N_K} \frac{\sum_j \sum_{X^i_{t-\tau}\in \mathcal P^j_k}  \mid \{{ \Phi}^j_k(i,\tau)\}^{reco.}  - \{\Phi^j_k(i,\tau)\}^{ref}\mid}{\sum_j |\mathcal P^j_k|}% \text{ for }k = 1, \dots, \{N_K\}^{ref}
\end{equation*}
Can be also computed as average \textit{percentage} error per regime:
\begin{eqnarray}%\label{eq:DPhirel_groundtruth}
&&{\Delta  \Phi} (\%) = \frac{1}{N_K} \sum_{k=1}^{N_K} \nonumber \\
&&\frac{\sum_j \sum_{X^i_{t-\tau}\in \mathcal P^j_k} \frac{ \mid\{{ \Phi}^j_k(i,\tau) \}^{reco.} - \{\Phi^j_k(i,\tau)\}^{ref} \mid }{\{\Phi^j_k(i,\tau)\}^{ref}}}{\sum_j |\mathcal P^j_k|} \times100  \nonumber   
\end{eqnarray}
% \subsubsection*{All tested links}
% Alternatively, one could compute the error by considering all the tested links. This estimate 'rewards' also having good FPR.\el{?}
% \begin{equation}\label{eq:DPhi_all}
%  {\Delta \Phi}_k = \frac{1}{N^2_{var}\cdot \tau_{\max}} \sum_{j=1}^{N} \sum_{i=1}^{N}\sum_{\tau=1}^{\tau_{\max}} [{\hat \Phi}^j_k(i,\tau)  - \{\Phi^j_k(i,\tau)\}^{ref}] %\text{  for }k = 1, \dots, \{N_K\}^{ref}
% \end{equation}

\subsection{Prediction error}
\begin{equation*} \label{eq:predition_err}
\hat \varepsilon \equiv \frac{1}{N_X T } \sum_t \sum_j|\{x^j(t)\}^{ref}- \{ x^j(t)\}^{reco.}|  \approx \sqrt{ \frac{\mathbf{L}}{N_{X}\cdot T}}
\end{equation*}
with $\mathbf{L}$ defined in Eq (\ref{eq:costfunctional_reg}).

\section{Abbreviations and notations}
\begin{table}[ht]%H
\begin{center}
\begin{tabular}{lll}
\multicolumn{2}{c}{Abbreviations}\\
\hline
%IC 	& 		&  Akaike  Information  criterion\\
AIC 	& 		&  Akaike  Information  criterion\\
AICc			&		&   Corrected  Akaike  Information  criterion\\
ENSO			&		&  El Ni\~no Southern Oscillation\\
FPR		&		&    False positive rate\\
MCI && Momentary conditional independence\\
% MLR 		&		&    multi linear regression\\
PCMCI 			& 		& Causal discovery method \cite{Runge2019b}\\
RAM			&		& Regime-dependent Autoregressive Model \\
SCM & & Structural causal model\\
TPR			&		&    True positive rate\\

\hline
\end{tabular}
\end{center}
\caption{Abbreviations used throughout the manuscript.}
\label{tab:Abbreviations}
\end{table}

\begin{table}[ht]%H
\begin{center}
\begin{tabular}{lll}
\multicolumn{2}{c}{List of notation}\\
\hline
$\{X_t\}_{t\in \mathbb{Z}}$			&		&  Stochastic process\\
$N_X$ 			& 		& Spatial dimension of $\{X_t\}$\\
$N_K$ 	& 		&  Number of regimes\\
$N_C$			&		&   Bound for switches of $\gamma_k(t)$ for each $k$\\
$N_Q$ 		&		&   Number of iteration steps \\
$N_A$			&		& Number of annealing steps\\
$N_R$ 		&		&   Number of realisations\\
$N_{\text{para}}$ 		&		&   Number of parameters\\
$\alpha$ 		&		& Link significance level \\
$\mathcal{P}^j_t$			&		&    Parents of component $X^j_t$ \\
$\Gamma(t)$		&		&    Regime assigning process\\
$\Phi_t $ 		&		&   Causal effect parameters, time dependent \\
$\Upsilon_k$		&		&    Collection of specific time steps, dependent on regime\\
$\mathbf{x}_t$		&		& Time series\\
\hline
\end{tabular}
\end{center}
\caption{Notation used throughout the manuscript.}
\label{tab:Notations}
\end{table}

\bibliographystyle{plain}
\bibliography{survey_paper}

\end{document}